\documentclass{article}
\usepackage{graphicx}			
\usepackage{booktabs}			
\usepackage{tabulary}			
\usepackage[T1]{fontenc}		
\usepackage[utf8]{inputenc}		
\usepackage{xpatch}
\usepackage{xcolor}				
\usepackage{listings}			
\usepackage[page,toc,title]{appendix}
\usepackage{minitoc}
\usepackage{enumitem}
\usepackage{hyperref}
\usepackage{pdflscape} 
\hypersetup{
   colorlinks=true,
   linkcolor=black,
   citecolor=black,
   filecolor=magenta
}
\usepackage{apacite}

\usepackage{multibib}
\newcites{SI}{Supplementary Information References}

\usepackage{amsthm}
\newtheorem{RQ}{RQ}

\usepackage{authblk}
\usepackage{float}
\usepackage{tabularx,ragged2e}
\usepackage{longtable}
\usepackage{array}
\usepackage{amsmath}  
\usepackage{pifont}
\newcolumntype{L}[1]{>{\raggedright\let\newline\\\arraybackslash\hspace{0pt}}m{#1}}

\title{\vspace{-4cm}Winning and losing with Artificial Intelligence: What public discourse about ChatGPT tells us about how societies make sense of technological change}
\author[1]{Adrian Rauchfleisch \thanks{adrian.rauchfleisch@gmail.com}}
\author[1]{Joshua Philip Suarez} 
\author[1]{Nikka Marie Sales}
\author[2]{Andreas Jungherr}

\affil[1]{Graduate Institute of Journalism, National Taiwan University}
\affil[2]{Institute for Political Science, University of Bamberg}

\date{July 2025}

\begin{document}
\doparttoc 
\faketableofcontents 

\maketitle

\begin{center}
Working Paper
\end{center}

\begin{center}
\textbf{Abstract}
\end{center}
\begin{quote}
Public product launches in Artificial Intelligence can serve as focusing events for collective attention, surfacing how societies react to technological change. Social media provide a window into the sensemaking around these events, surfacing hopes and fears and showing who chooses to engage in the discourse and when. We demonstrate that public sensemaking about AI is shaped by economic interests and cultural values of those involved. We analyze 3.8 million tweets posted by 1.6 million users across 117 countries in response to the public launch of ChatGPT in 2022. Our analysis shows how economic self-interest, proxied by occupational skill types in writing, programming, and mathematics, and national cultural orientations, as measured by Hofstede’s individualism, uncertainty avoidance, and power distance dimensions, shape who speaks, when they speak, and their stance towards ChatGPT. Roles requiring more technical skills, such as programming and mathematics, tend to engage earlier and express more positive stances, whereas writing-centric occupations join later with greater skepticism. At the cultural level, individualism predicts both earlier engagement and a more negative stance, and uncertainty avoidance reduces the prevalence of positive stances but does not delay when users first engage with ChatGPT. Aggregate sentiment trends mask the dynamics observed in our study. The shift toward a more critical stance towards ChatGPT over time stems primarily from the entry of more skeptical voices rather than a change of heart among early adopters. Our findings underscore the importance of both the occupational background and cultural context in understanding public reactions to AI.
\end{quote}

\section{Introduction}
Technological change introduces friction in society \cite{frey_technology_2020}. While some actors capitalize on emerging opportunities, others face the obsolescence of their skills. Although the long-term outcomes of ongoing and future technological change remain uncertain, specific events can crystallize these processes and serve as focal points for public meaning-making \cite{birkland_after_1997, birkland_focusing_1998}. Severe technological accidents highlight risks, while impressive technological feats showcase potential. Crucially, it is not only the events themselves but the public negotiation of meaning \cite{dayan_media_1992,gamson_media_1989, lang_television_1984}\textemdash among pundits, stakeholders, and the broader public\textemdash that shapes discourse and influences societal perceptions of a technology’s benefits and risks. The launch of new technologies is often strategically staged as a media event designed to trigger widespread public engagement, media coverage, and discursive negotiation. When successful, they can become focusing events that shape the collective understanding of the underlying technology.

Artificial Intelligence (AI) is one such technology, surrounded by bifurcated societal expectations \cite{oshaughnessy_what_2023, zhang_artificial_2019, bullock_public_2023}. Some view AI as a tool to empower workers and enhance productivity, while others fear it will eliminate jobs and render many workers obsolete. Despite this ongoing debate, it is essential to acknowledge that AI remains an abstract and often opaque technology for much of the public. Many people lack detailed knowledge of its workings, have not yet directly interacted with it, and are only beginning to encounter its integration into everyday life. In this context, focusing events\textemdash such as high-profile AI product launches, critical failures, or surprising successes\textemdash play a central role in triggering collective meaning-making processes \cite{jungherr_artificial_2024}. Studying public reactions to such events can yield important insights into how societies interpret and adapt to technological change.

In this article, we examine the public product launch of ChatGPT and the subsequent public negotiation of its meaning on Twitter (now X). We are particularly interested in how individuals’ expected economic interests and cultural orientations shape their engagement with the debate. Specifically, we ask: Who engages in public discourse following the launch of ChatGPT, when do they engage, and how do they frame the technology?

ChatGPT is widely seen as a powerful tool for automating cognitive labor, especially in the cultural sector. We therefore expect individuals' economic self-interest, as proxied by their occupational skill level, to be associated with both (i) their likelihood and timing of engagement (measured by the speed of their response to the event), and (ii) their stance toward the technology. Because ChatGPT complements tasks such as programming and data analysis, individuals in roles relying on these tasks are expected to engage early and positively, while those whose work tasks rely primarily on writing or content production skills see the technology less than a complement and more as a replacement of their skills and may join later and with greater skepticism. This lag is attributed to differences in subjective topical affinity with emergent technologies as well as diverging views on the likely impact of generative AI on one’s own economic fortunes.

In addition, we examine the role of national cultural context. ChatGPT exemplifies a novel, and partly uncertain, technological innovation, and prior research suggests that societal openness to innovation varies systematically across cultures. To operationalize cultural variation, we draw on \citeauthor{hofstede_dimensionalizing_2011}'s \citeyear{hofstede_cultures_2010, hofstede_dimensionalizing_2011} well-established framework of cultural dimensions. Following Hofstede’s original dimensions and argument, we assume that individuals embedded in cultures characterized by low uncertainty avoidance, high individualism, and low power distance will (i) be more likely to engage early with the topic and (ii) express more positive views of the technology. 

To test the roles of economic interest and culture, we analyzed Twitter discourse on ChatGPT from November 30, 2022, to February 1, 2023. We analyzed the stance towards ChatGPT in tweets (n = 3,793,601 tweets) and identified the professional background and country of origin of users (n = 1,606,270 users). Our analysis reveals that individuals engage in the public negotiation of meaning surrounding ChatGPT, aligning with their expected economic interests and cultural backgrounds. Importantly, aggregate trends in public sentiment obscure these individual-level dynamics. In the case of ChatGPT, the overall discussion did not become more negative due to a general shift in opinion, but rather due to a shift in the composition of participants. Early discourse was dominated by technologically optimistic individuals who were likely to benefit from AI, as well as by actors embedded in more individualistic cultural contexts. As the discussion broadened, it included more skeptical voices\textemdash often those with less to gain or more to lose, and from more cautious cultural contexts. Thus, changes in aggregate sentiment reflect changes in speaker composition rather than a genuine shift in attitudes toward the technology.

\section{Theory}
\subsection{Focusing Events and the Collective Negotiation of Meaning}
In times of uncertainty, societies engage in collective sensemaking—a process that often includes the public negotiation of meaning in response to focusing events. These events, whether sudden or staged, draw attention to the potential power and consequences of emerging technologies. Technological change, in particular, has repeatedly prompted such negotiations, especially around the perceived benefits and risks of innovation \cite{gamson_media_1989}.

The effects of new technologies are frequently abstract and removed from everyday experience. Most people lack direct knowledge or experience with these innovations, making mediated public discourse a crucial mechanism through which individuals come to understand and evaluate technological developments, such as artificial intelligence (AI). Events play an outsized role in this discourse, serving as focal points for collective meaning-making. These may include naturally occurring incidents, such as accidents or system failures, or carefully staged moments, such as press conferences or product launches \cite{birkland_after_1997, birkland_focusing_1998}. Crucially, these events trigger extensive media coverage and spur communicative activity among politicians, experts, and citizens, generating interpretations and contested framings \cite{dayan_media_1992, lang_television_1984}.

The purpose and function of these negotiations differ depending on the theoretical perspective. From a rationalist viewpoint, public discourse is a deliberative process aimed at exchanging arguments, identifying shared concerns, and developing consensual interpretations and prospective solutions \cite{habermas_faktizitat_1998}. In contrast, communitarian and cultural perspectives emphasize the identity-forming aspect of such episodes, framing them as moments of symbolic cohesion and collective belonging \cite{anderson_imagined_2016, dayan_media_1992, lang_television_1984}. Conflict-oriented and power-analytical approaches highlight yet another role: focusing events as arenas for strategic position-taking by societal actors who have material or symbolic stakes in technological change. In this view, public discourse rarely settles into shared positions. Instead, it shifts depending on the nature of the event, its alignment with specific interests, and which actors are mobilized into the discursive space at a given moment \cite{gamson_media_1989}.

Over the past two decades, product launches and public beta releases have become central elements in the strategic communication repertoire of technology firms. Apple’s keynote events, notably under Steve Jobs, exemplify the performative presentation of innovation and the introduction of products designed to trigger public interest, interaction, and discussion \cite{wenzel_strategy_2018}. Other technology companies have adopted and adapted this format. In the recent wave of AI innovations, carefully staged product launches by firms such as OpenAI and Nvidia play a pivotal role in shaping public narratives, setting expectations, and influencing evaluations of companies and technologies. These events thus serve as important instances of public sensemaking, offering valuable opportunities to study the societal negotiation of AI’s meaning by interested and affected actors.

The rise of digital media has significantly improved our ability to observe and analyze these processes. By tracing online discourse across platforms, we can investigate who participates in public debates following such events, when they engage, how long they remain active, and what arguments they put forward. Among the most promising sources of data for such analysis is the microblogging service Twitter. As a platform frequented by technology professionals, media actors, consultants, and academics, Twitter plays a central role in real-time discursive engagement. In this context, the platform’s demographic bias—often seen as a limitation—becomes an analytical strength, offering insight into digitally literate and professionally invested publics. Prior research has demonstrated Twitter’s value for capturing collective sensemaking processes, for example, in response to televised election debates \cite{jungherr_logic_2014} or high-profile product launches such as Apple keynotes \cite{lipizzi_towards_2016}.

Building on these insights, we examine the public launch of ChatGPT as a focusing event that triggered widespread discourse on Twitter. We conceptualize this moment as part of a broader societal negotiation over the meaning and implications of AI. Specifically, we analyze how individuals’ occupational background (as a proxy for economic self-interest) and cultural orientation shape both their engagement with and interpretation of ChatGPT. Our central questions are: Who speaks when, and how do their positions reflect their economic and cultural embeddedness?

\subsection{The ChatGPT launch as a focusing event}
ChatGPT is an AI-powered chatbot launched publicly on November 30, 2022, that responds to user queries using GPT-3, a pretrained autoregressive language model developed by OpenAI \cite{brown_language_2020}. The discussions surrounding OpenAI’s chatbot ChatGPT \cite{ouyang_training_2022} offer a valuable window into the collective negotiation of meaning and public sensemaking regarding the societal, political, and economic implications of AI-powered tools and services. Its launch was consciously designed to showcase the capabilities of transformer models through a user-friendly interface, enabling anyone to generate meaningful content via simple text prompts. This made the application of AI tangible and experiential for a broad public. Typically, public debates about AI remain abstract and disconnected from people’s daily experiences. In contrast, ChatGPT allowed individuals to interact directly with cutting-edge AI and form opinions based on first-hand use.

Prior to ChatGPT, other public-facing prototypes, such as OpenAI’s DALL·E, which generates images from text prompts \cite{ramesh_hierarchical_2022}, had also garnered attention. However, these generated considerably less public discussion. There are at least three likely reasons for ChatGPT’s broader resonance. First, text production is more central to a wider range of occupations and tasks than image production. Consequently, advances in this area are relevant to a broader audience and impact more sectors of the economy. Moreover, text is a more ubiquitous medium in digitally mediated communication, making the automation of text generation more visible and broadly applicable than automated image generation. Second, AI-generated text is often more difficult to distinguish from human-written text than AI-generated images are from those created by humans. This ambiguity heightens both the perceived power and relevance of text-generating models. Third, ChatGPT's public impact was amplified by OpenAI’s strategic partnership with Microsoft, which integrated ChatGPT functionality into widely used products. This business endorsement added institutional legitimacy and spurred additional public and media attention. Taken together, these factors make ChatGPT and the discourse it generated a compelling case for studying public narratives surrounding the social, economic, and political implications of AI.

The launch of ChatGPT functioned as a media event in the classical sense, as described by \citeA{dayan_media_1992}: a carefully staged, high-visibility occasion designed to attract and coordinate public attention, frame collective experience, and foster shared cultural engagement. Rather than unfolding organically or remaining confined to expert communities, OpenAI’s launch was a ritualized moment of publicity, engineered to break routine, create spectacle, and invite mass participation. In contrast to traditional AI model releases through technical white papers or API-limited access, ChatGPT was introduced via a publicly accessible web interface, supported by coordinated announcements across blogs, news media, and especially social platforms. This strategy aligned with what \citeA{couldry_media_2003} refers to as the “media rituals” through which institutions assert authority and meaning via performance in mediated space.

Crucially, the launch was not only a communication act but also a performance of innovation, a “showcase” that activated sociotechnical imaginaries \cite{jasanoff_containing_2009} about the future of AI in everyday life. OpenAI invited users to “try out” ChatGPT for free, democratizing access while staging a participatory demonstration of technological capability. Millions of users engaged with the chatbot within days, generating and sharing outputs that highlighted both ChatGPT’s remarkable fluency and its amusing or troubling failures. This user-generated discourse, especially on Twitter, did not merely document the product; it enacted a public negotiation of meaning, transforming individual interactions into a broader cultural event. As such, the launch bridged the gap between technical development and social discourse, transforming private experimentation into collective sensemaking.

OpenAI’s communicative choreography thus blurred the lines between product launch, public beta, media spectacle, and cultural script. It was not merely the release of a tool but the construction of a moment in which technological potential was made tangible, discussable, and debatable. The strategic openness of the launch amplified its resonance, allowing the public to project hopes, fears, and values onto the technology. In this sense, the event exemplifies the mediatization of innovation \cite{hepp_deep_2020}, where technological change is increasingly introduced, interpreted, and legitimized through media practices. As a media event, the launch of ChatGPT did not just inform audiences; it invited them to imagine, assess, and contest the societal role of AI, offering a fertile context for studying how actors with different interests, identities, and cultural backgrounds negotiate the meaning of emerging technologies in real-time.

A growing body of research has examined how users on Twitter responded to ChatGPT’s release. However, most of this work remains descriptive and aggregate in nature. Many early studies focus on sentiment analysis, classifying tweets as positive, neutral, or negative \cite{futterer_chatgpt_2023, haque_i_2022, koonchanok_public_2024, leiter_chatgpt_2024, miyazaki_public_2023, nasayreh_arabic_2024, weber_social_2024, xing_voices_2024}. These studies consistently show a generally positive or neutral sentiment in the months following the release. In domain-specific analyses, for example, in the field of education, researchers find that while early responses were primarily positive, sentiment diversified throughout 2023, with some spikes in negativity on specific days \cite{futterer_chatgpt_2023}. Nonetheless, overall negative sentiment declined over time \cite{koonchanok_public_2024}.

Topic modeling analyses provide further insight into evolving themes. For instance, Japanese users initially emphasized future technological possibilities, while later discussions shifted toward concrete applications \cite{fukuma_comparative_2024}. One early study found that tweets referencing careers and software development were, on average, more positive than those referencing education \cite{haque_i_2022}.

While these studies offer valuable insights into aggregate trends and thematic developments, they largely overlook who participates in these discussions and why. Few studies consider the role of occupational background or skill level in shaping engagement and attitudes. Similarly, cross-cultural comparisons are rare. Most existing research maps the “what” of public discourse. However, it leaves the “who” and “why” underexplored, questions we aim to address by examining the relationship between occupation, culture, and responses to ChatGPT.

\subsection{Explaining people’s positions}
Examining the positions people take in the public negotiation of meaning reveals two key factors that shape their responses at both the group and systemic levels: the economic interests they hold and the cultural values in which they are embedded.

\subsubsection{The role of economic interest}
We expect individuals’ attitudes toward ChatGPT to be shaped by their economic self-interest. When assessing prospective change, people are likely to interpret it through the lens of their own role in the anticipated future. This is especially true for technological change. New technologies—particularly those that enable the automation of tasks—create opportunities for some while threatening the livelihoods of others. Individuals with skills that align well with the new technology tend to benefit, while those whose skills are more easily automated may face diminished prospects.

In the case of ChatGPT, we expect automation to particularly affect text-based tasks, or more broadly, tasks involving symbolic manipulation. This may benefit individuals with skills in areas such as computer programming or mathematics, as these are instrumental in developing and applying AI systems. By contrast, large language models (LLMs) may automate aspects of technical writing, marketing, and teaching, potentially threatening those professions.

We anticipate that these differing economic stakes will shape how people engage in the public negotiation of ChatGPT’s meaning and role. Those likely to benefit are expected to speak positively about the technology and to engage early in public discourse. Conversely, those whose skills are more at risk of being automated may be more critical and engage later in the conversation.

Few studies have explicitly focused on how an individual's occupational background influences their reactions to ChatGPT on Twitter. \citeA{koonchanok_public_2024} match job titles in the O*NET database with Twitter users’ descriptions and calculate sentiment across 23 broad categories, finding that “research” roles report slightly more negative views than “sales,” yet overall variance remains limited. \citeA{miyazaki_public_2023} also identify occupations based on users’ descriptions by using two different dictionaries from O*NET and the job platform Indeed. They then combine specific occupations with sentiment analysis and show that, for example, data scientists or product managers express a significantly more positive sentiment toward generative AI than illustrators or writers. This finding contrasts with their expectation, as they had assumed that occupation groups at risk of job replacement through AI would react more negatively. \citeA{giordano_impact_2024} push this further by extracting tasks from tweets and mapping them to 185 European Classification of Skills, Competences, and Occupations (ESCO) skills, revealing that writing‐related skills are among the most polarized. These studies demonstrate variation in reactions based on occupational background and the skills most commonly discussed in the context of ChatGPT. However, none of them combine both and specifically predict the sentiment towards ChatGPT.

One potential explanation for attitudes toward ChatGPT could be the risk of job replacement \cite{miyazaki_public_2023}. Prior research has shown that specific jobs have a higher risk of being replaced through computerization \cite{frey_future_2017}, which also includes AI, or specifically, through AI \cite{acemoglu_artificial_2022}. While these earlier studies use broader approaches to estimate long-term job replacement, we argue that ChatGPT, with its main focus on text and chat, primarily challenges writing, which is, as prior studies have indicated, a central element in the Twitter discourse \cite{giordano_impact_2024, koonchanok_public_2024}. We also assume, based on the prominence and positive sentiment surrounding programming and other technical themes in ChatGPT discourse \cite{haque_i_2022}, that programming and mathematics may potentially influence attitudes toward ChatGPT. For all three skills, the association could go in both directions, as ChatGPT, with its capabilities, can be seen as either a threat or a helpful tool for individuals with these skills. Supporting this assumption, \citeA{miyazaki_public_2023} show that sentiment varies considerably across occupations, with more positive views often expressed by those whose work aligns more closely with ChatGPT’s capabilities, such as data scientists or product managers, while illustrators and writers, whose core tasks are more directly challenged, tend to express more concern.
\begin{RQ}How are occupational skills associated with attitudes toward ChatGPT?\end{RQ}
Although prior research has not explicitly examined the timing of first engagement with ChatGPT in relation to occupational background, it is clear that early discussions predominantly focused on the technological potential of ChatGPT \cite{haque_i_2022}. Over time, however, these discussions became increasingly diverse in their topics and shifted toward more critical assessments of the technology’s limitations, particularly among late adopters \cite{fukuma_comparative_2024}. According to \citeA{rogers_diffusion_1983}, early adopters of any technology tend to exhibit distinct characteristics compared to late adopters, including differences in their social networks and openness to innovations. Building on this, we aim to investigate whether the timing of ChatGPT adoption, the first time people started to engage with it, is influenced by occupational background and related skills.

\begin{RQ}How are occupational skills associated with the time people adopted ChatGPT for the first time?\end{RQ}

While occupations vary within countries, cultural orientations operate across them. Together, these two factors allow us to model both group-level and system-level influences on attitudes toward ChatGPT.

\subsubsection{The role of culture}
People are also embedded in cultural contexts. Following \citeA{hofstede_cultures_2010}, we understand culture as the shared values, norms, and behaviors learned within a group, which shape how individuals perceive, think, and act in social settings. These cultural frameworks are likely to influence how people evaluate the societal impact of ChatGPT. Individuals from the same cultural background are, therefore, more likely to share similar views. In the context of AI, several cultural dimensions are particularly relevant: uncertainty avoidance, individualism, and egalitarianism.

Culture has been largely absent in studies analyzing reactions to ChatGPT on Twitter. While some prior studies have reported the geographic distributions of tweets and users \cite{futterer_chatgpt_2023, haque_i_2022}, the role of culture has not been the focus of these studies. Some studies specifically analyze subsets of Twitter users, such as those writing in Arabic \cite{mujahid_arabic_2023} or Japanese \cite{fukuma_comparative_2024}. However, they do not explicitly discuss cultural factors and differences. Only \citeA{miyazaki_public_2023} briefly mention the potential role of culture in their discussion of future research. Another exception is \citeauthor{leiter_chatgpt_2024}'s \citeyear{leiter_chatgpt_2024} study, which shows that sentiment is more negative in languages other than English. They also find small differences regarding issue prevalence across languages. In German, for instance, education and social concerns were more prevalent than in English. Both findings could be indicators of cultural differences. The most detailed study in this context is Murayama et al.’s (2025) \citeauthor{murayama_linguistic_2025}'s \citeyear{murayama_linguistic_2025} analysis of 14 different languages on Twitter, which reveals differences across language communities in relative interest and sentiment about AI, but does not systematically test or operationalize cultural factors.

Therefore, our study focuses on the potential system-level factor of culture. While cultural variables can be measured individually \cite{schwartz_measuring_2022, swedlow_construct_2020}, \citeA{hofstede_dimensionalizing_2011} defines culture as a collective phenomenon that should be operationalized only at the system level. For Twitter users, we only have access to system-level group membership (countries), and thus \citeauthor{hofstede_dimensionalizing_2011}'s \citeyear{hofstede_dimensionalizing_2011} six dimensions of culture could be considered for the analysis. Similar to the Twitter studies, culture has been understudied in the context of AI. 

An exception is \citeauthor{wilczek_government_2024}'s \citeyear{wilczek_government_2024} study, which focuses on support for AI regulation and uses Hofstede’s uncertainty-avoidance dimension, capturing how comfortable people feel with ambiguity and uncertainty as the main predictor for regulation preferences. While the connection is primarily between regulation in general and uncertainty avoidance, there is also a potential link to technology adoption, as prior research comparing the US (a medium-low) and Germany (a medium-high uncertainty-avoidance culture) has shown \cite{eitle_cultural_2020}. Therefore, there is a strong argument that people from cultures with high uncertainty avoidance will have less positive sentiment toward ChatGPT.

Individualism is another cultural variable that has been identified in the context of cultural cognition. This dimension, spanning from individualism to communitarianism, is predictive of more positive attitudes toward AI \cite{oshaughnessy_what_2023}. Therefore, we expect individualism to be associated with more positive attitudes toward ChatGPT. 

The third cultural variable on which we base our expectations is egalitarianism, which is closely related to \citeauthor{hofstede_dimensionalizing_2011}'s \citeyear{hofstede_dimensionalizing_2011} power distance dimension. Egalitarians, thus low in power distance, perceive more benefits from AI \cite{oshaughnessy_what_2023}. We do not include the remaining three dimensions from Hofstede’s framework in our study, as we do not have expectations for their relevance.

\begin{RQ}How are cultural factors associated with attitudes toward ChatGPT?\end{RQ}

We are also interested in how cultural factors explain when people first engage with new technology. Prior research on technology acceptance has shown that higher individualism correlates with greater intentions to use technology, while higher power distance is associated with a greater likelihood of adopting technology \cite{jan_hofstedes_2024}. Uncertainty avoidance \cite{wilczek_government_2024} is also a reasonable predictor in the context of AI adoption. Given these findings, we expect that some of these broader trends from the technology acceptance literature may also apply to the adoption of ChatGPT.

\begin{RQ}How are cultural factors associated with the time people adopted ChatGPT for the first time?\end{RQ}

By focusing on these four research questions, we aim to show that public sentiment shifts not because of a change in opinion but due to the evolving composition of those engaging in the discussion.

\section{Data and methods}
To answer our research questions, we rely on Twitter data. While the Twitter population is not representative of a country's population, the biased nature of Twitter users\textemdash namely, a strong prevalence among people belonging to publics interested or invested in digital technology, consultants, public intellectuals, trainers, and academics\textemdash is, for once, a strength of running analyses on messages posted on this microblogging service. On Twitter, we find an important slice of the societal negotiation of meaning among invested individuals and publics. In prior work, Twitter data has served as windows into similar collective negotiations of meaning and public sensemaking, as in the case of televised debates during election campaigns \cite{jungherr_logic_2014}. We used Twitter’s Stream and Academic API to collect ChatGPT-related tweets with the search terms “chatgpt” and “gpt”. For the first few days after the launch of ChatGPT, we searched for tweets, and starting from 7 December 2022, we directly captured tweets over the stream API. We then retained only tweets that mentioned ChatGPT (“chat-gpt”, “chatgpt”, or “chat gpt”), resulting in a final sample of 3,793,601 tweets written by 1,606,270 unique users from November 30, 2022, to February 1, 2023 (UTC). 

To answer our research questions, three aspects require classification: (i) stance of the tweet toward ChatGPT (ii) country of the user posting the tweet, and (iii) occupation of the user posting the tweet.  For all three classification tasks, we relied on OpenAI’s GPT-4o mini model (“gpt-4o-mini-2024-07-18”).

\subsection{Stance detection}
For the stance detection, we also relied on ChatGPT (for information about the classification and manual validation, see \hyperref[app:SI_stance]{Supplementary Information A.1}). We first tested several off-the-shelf sentiment classifiers, but they did not perform well when we tried to validate them. Another drawback was that these traditional approaches typically struggled with multilingual text classification. Prior research \cite{rathje_gpt_2024} demonstrates that GPT models surpass other methods in multilingual classification tasks. The classifier worked well, as our manual validation indicates (accuracy=0.87; Cohen’s $\kappa$=0.78). Tweets were classified as either communicating a negative, neutral, or positive stance towards ChatGPT.

\subsection{Location identification}
We relied on the location field provided by the users in their Twitter profiles to identify the country from which the tweets were posted. We first tested OpenStreetMap for the location identification but then opted for ChatGPT to identify the location of users because of substantially better performance (for more information, see \hyperref[app:SI_loc]{Supplementary Information A.2}). 1,080,332 unique users provided information in their location field. With our classification approach, we identified the country for 715,890 unique users. We post-processed the identification results and manually cleaned systematic errors. The results were validated through a 500 random sample of users with identified locations and another 500 random sample of users with locations not identified (NA cases) using the method previously described. Manual validation results showed high accuracy for both users with identified locations (96.4\%) and without an identified location (95.6\%); in both cases, we only verified whether the assigned country correctly matched the user’s self-reported location field. We then carried out an in-depth validation by comprehensively reviewing each sampled user’s full profile, beyond just the location field, to confirm their true country of origin. This holistic approach yielded 82.65\% accuracy in reasonable country assignments. We then matched each inferred user (and their tweets) to per-country data for Hofstede’s cultural dimensions \cite{hofstede_cultures_2010, the_culture_factor_group_country_nodate}. A total of 1,739,212 tweets or 704,926 unique users have information about Hofstede’s cultural measure.

\subsection{Occupation identification}
Prior research has already matched jobs based on dictionaries from O*NET lists \cite{miyazaki_public_2023} or by directly matching job titles used in the O*NET database with jobs mentioned in user descriptions \cite{koonchanok_public_2024}. We first tested whether GPT-4o mini has access to O*NET's standardized jobs and can correctly connect user descriptions with job IDs. After manually validating this approach, we classified all users based on their description (for more details about the classification and validation, see \hyperref[app:SI_occ]{Supplementary Information A.3}). In total, 493,946 unique users with 1,404,782 tweets could be classified. 

\subsection{Empirical strategy}
We examine the influence of occupational skills and cultural factors across two main outcomes: (i) the attitude toward ChatGPT, and (ii) the timing of initial engagement with ChatGPT discourse. To account for clustering effects in the data structure, we use multilevel models with random intercepts for both occupation and country. The independent variables include: writing, programming, and mathematics skill levels (from O*NET), as well as power distance index (PDI), individualism versus collectivism (IDV), and uncertainty avoidance index (UAI) from Hofstede’s cultural dimensions. Parameters are estimated using maximum likelihood estimation.

First, we estimate a multilevel multinomial logistic regression model to analyze the factors associated with a tweet's stance toward ChatGPT (for RQ1 and RQ3). The dependent variable is categorical (positive, neutral, negative), with neutral as the reference category. We conduct analyses using the mclogit package in R. 

Second, we develop a multilevel Cox proportional hazards model to analyze the time until a user first engages with ChatGPT discourse (for RQ2 and RQ4). The observation window begins with the launch of ChatGPT (November 30, 2022, 6:00 PM UTC), and the event of interest is a user’s first tweet related to ChatGPT. The outcome is the elapsed time (in seconds) from the launch until first engagement. We conduct analyses using the \textit{coxme} package in R.

Additionally, we conducted several robustness checks. First, for each model, we separately estimate models including only (i) occupation-level factors (N tweets = 1,404,657; N users=493,946) and (ii) country-level factors (N tweets = 1,739,101; N users = 704,926), as it allowed us to include more observations. Second, we re-estimate models using skill importance (rather than skill level) from O*NET. Third, for stance models (addressing RQ1 and RQ3), we also estimate (i) a multilevel ordinal logistic model treating the stance of the tweet as an ordered outcome (negative as -1, neutral as 0, positive as 1) and (ii) a multilevel linear regression model with the user’s average stance score as a continuous dependent variable. The full results of the robustness tests are presented in the \hyperref[app:SI_rob]{Supplementary Information D}. They all support the findings that we report in the following section.

\section{Results}
We present our findings in three main parts. First, we provide a brief overview of the general trends in our dataset. Second, we examine how occupational skills affect users’ attitudes toward ChatGPT (RQ1) and the timing of their initial engagement (RQ2). Third, we discuss how cultural factors influence both sentiment (RQ3) and timing of engagement (RQ4). The full statistical models, along with visualizations of the conditional effects, are available in \hyperref[app:SI_full]{Supplementary Information C}.

\subsection{General descriptive analysis}
\begin{figure}[!htb]
\centering
\includegraphics[width=\textwidth,height=\textheight,keepaspectratio]{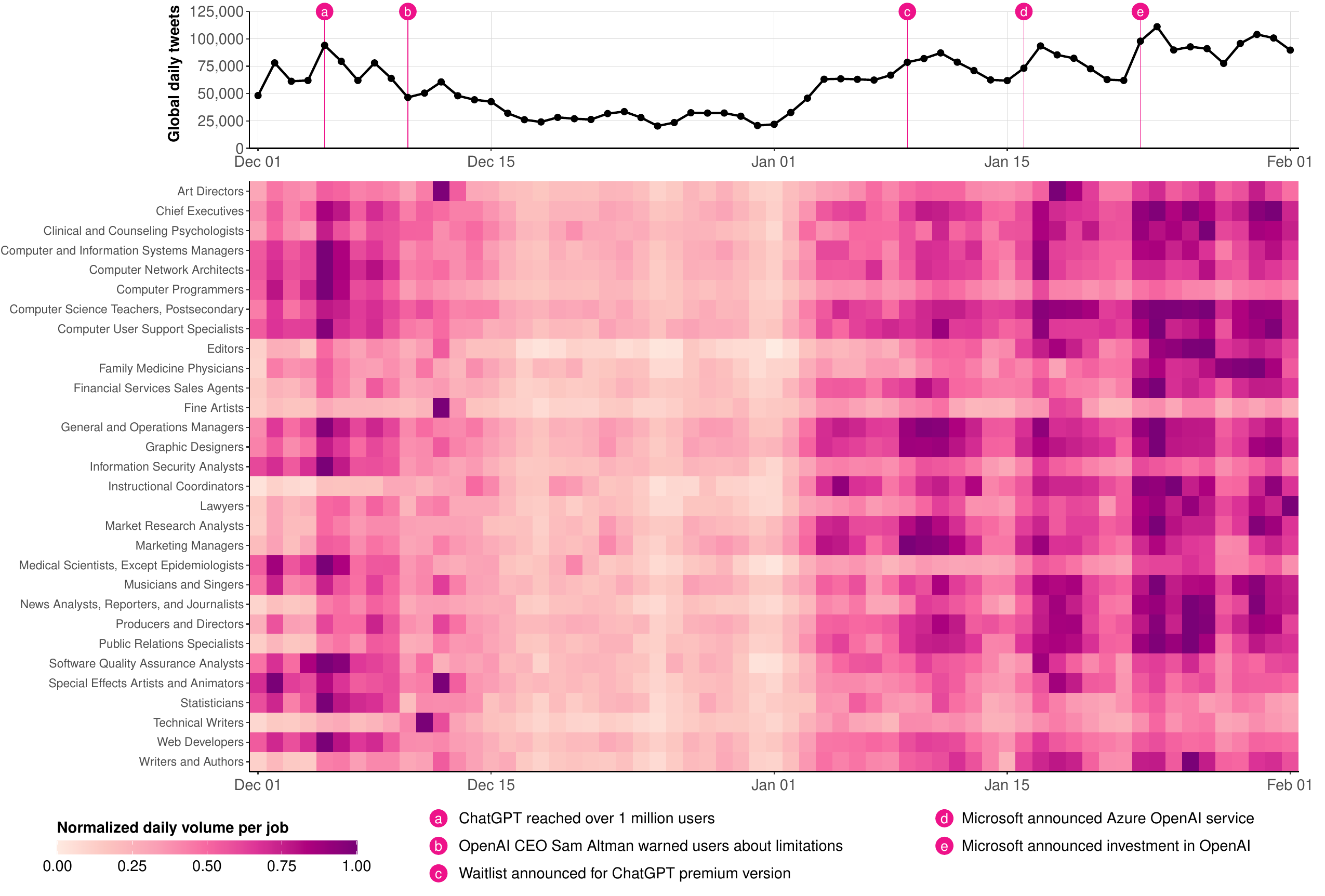}
\caption{Temporal distribution of tweets across occupations. Only the 30 occupations with the highest tweet volumes are shown.}
\label{fig:fig_01}
\end{figure}

Our dataset reveals strong initial interest among Twitter users in the early days following ChatGPT’s launch, particularly as users began exploring the platform’s capabilities. About 20\% of sampled users joined the conversation within the first 12 days (Figure \ref{fig:fig_02} and Figure \ref{fig:fig_04}). Tweet levels relatively declined around mid-December until the new year, but increased again in January, following announcements about Microsoft’s interest and investment, as well as the introduction of ChatGPT premium services (Figure \ref{fig:fig_01} and Figure \ref{fig:fig_03}). The sentiment of tweets throughout the study period was mostly neutral (51.0\%) or positive (30.6\%). However, certain issues occasionally triggered increases in negative sentiment, particularly those affecting specific occupations or cultures (see Figure \ref{fig:fig_02_app} and Figure \ref{fig:fig_03_app} in Supplementary Information B). In terms of language, most tweets in our sample were in English (68.0\%), followed by Japanese (12.7\%), French (4.3\%), and Spanish (4.3\%) (see Figure \ref{fig:fig_01_app} in Supplementary Information B).

\subsection{RQ1 and RQ2: Skills and occupations}
 The occupations that mainly contributed to ChatGPT discourse were computer programmers (14.3\%), general and operations managers (11.2\%), and graphic designers (8.7\%), accounting for the highest tweet volumes (see Figure \ref{fig:fig_01_app} (Panel A) in Supplementary Information B).

Regarding attitudes (RQ1), we found that users in occupations requiring higher levels of mathematics skills were more likely to post positive messages (Coef. = 0.087; 95\% CI = [0.045, 0.128]; p < 0.001) and less likely to post negative messages (Coef. = -0.060; 95\% CI = [-0.104, - 0.015]; p = 0.008), relative to neutral content. Meanwhile, higher writing skill requirements were associated with a lower likelihood of posting positive tweets (Coef. = -0.104; 95\% CI = [-0.150, -0.059]; p < 0.001). This supports earlier assumptions that occupations whose main tasks overlap substantially with ChatGPT’s capabilities, particularly in writing and content creation, may perceive the new technology as a threat. This was evident, for instance, following a widely circulated tweet on December 10, 2022, in which a user described successfully publishing a children’s book using ChatGPT and other AI tools. The tweet elicited a strong wave of negative responses, particularly among writers, authors, editors, art directors, and other creative professionals (see Supplementary Information B Figure \ref{fig:fig_02_app}), who raised concerns regarding ethics, artistic integrity, and the potential devaluation of creative labor (see \hyperref[app:SI_example]{Supplementary Information E} for example tweets).

\begin{figure}[!htb]
\centering
\includegraphics[width=\textwidth,height=\textheight,keepaspectratio]{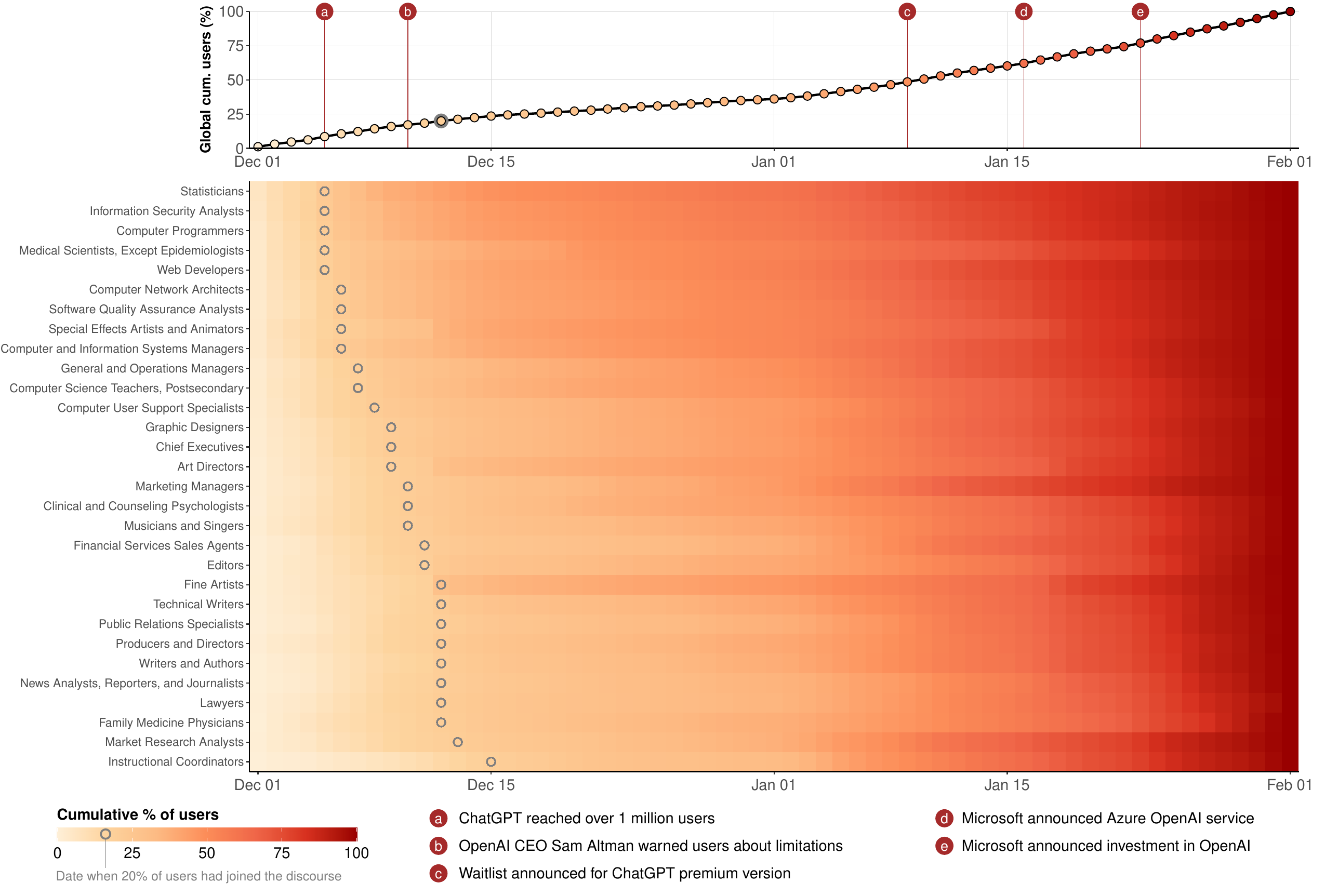}
\caption{Timing of initial engagement with ChatGPT discourse across occupations. Only the 30 occupations with the highest tweet volumes are shown. Occupations are ordered by the date at which 20\% of users first engaged with the discourse.}
\label{fig:fig_02}
\end{figure}

In terms of the timing of initial engagement (RQ2), we found that users whose occupations involve higher levels of programming skills were significantly more likely to engage earlier in the ChatGPT discourse (Coef. = 0.095; 95\% CI = [0.078, 0.112]; p < 0.001), all else being equal. This is expected, as programming-related users are naturally inclined to test new technologies early, especially ones that provide coding support such as ChatGPT. Statisticians, information security analysts, computer programmers, and web developers were among the earliest to engage (see Figure \ref{fig:fig_02}), particularly during the initial surge when ChatGPT reached one million users. In contrast, users in occupations requiring higher levels of writing skills were less likely to engage early (Coef. = -0.027; 95\% CI = [-0.047, -0.006]; p = 0.010), ceteris paribus. Technical writers, writers, authors, and editors generally exhibited later engagement patterns.

\subsection{RQ3 and RQ4: Cultural factors}
Regarding cultural background, the majority of tweets were generated by users from the United States (26.3\%), followed by India (7.8\%), Japan (6.5\%), the United Kingdom (6.5\%), and France (5.4\%) (see Figure  \ref{fig:fig_01_app} - Panel B in Supplementary Information). In terms of stance (RQ3), users from individualistic countries were less likely to express positive sentiment (Coef. = -0.003; 95\% CI = [-0.004, -0.001]; p = 0.005) and more likely to express negative sentiment (Coef. = 0.006, 95\% CI = [0.004, 0.007]; p < 0.001), relative to neutral messages. One potential explanation for this is that users in individual cultures might perceive ChatGPT primarily as a threat to personal autonomy, creativity, or job security. In contrast, users from collectivist cultures may focus more on potential communal or societal benefits, where collective advantages may outweigh personal concerns about disruption or change.

\begin{figure}[!htb]
\centering
\includegraphics[width=\textwidth,height=\textheight,keepaspectratio]{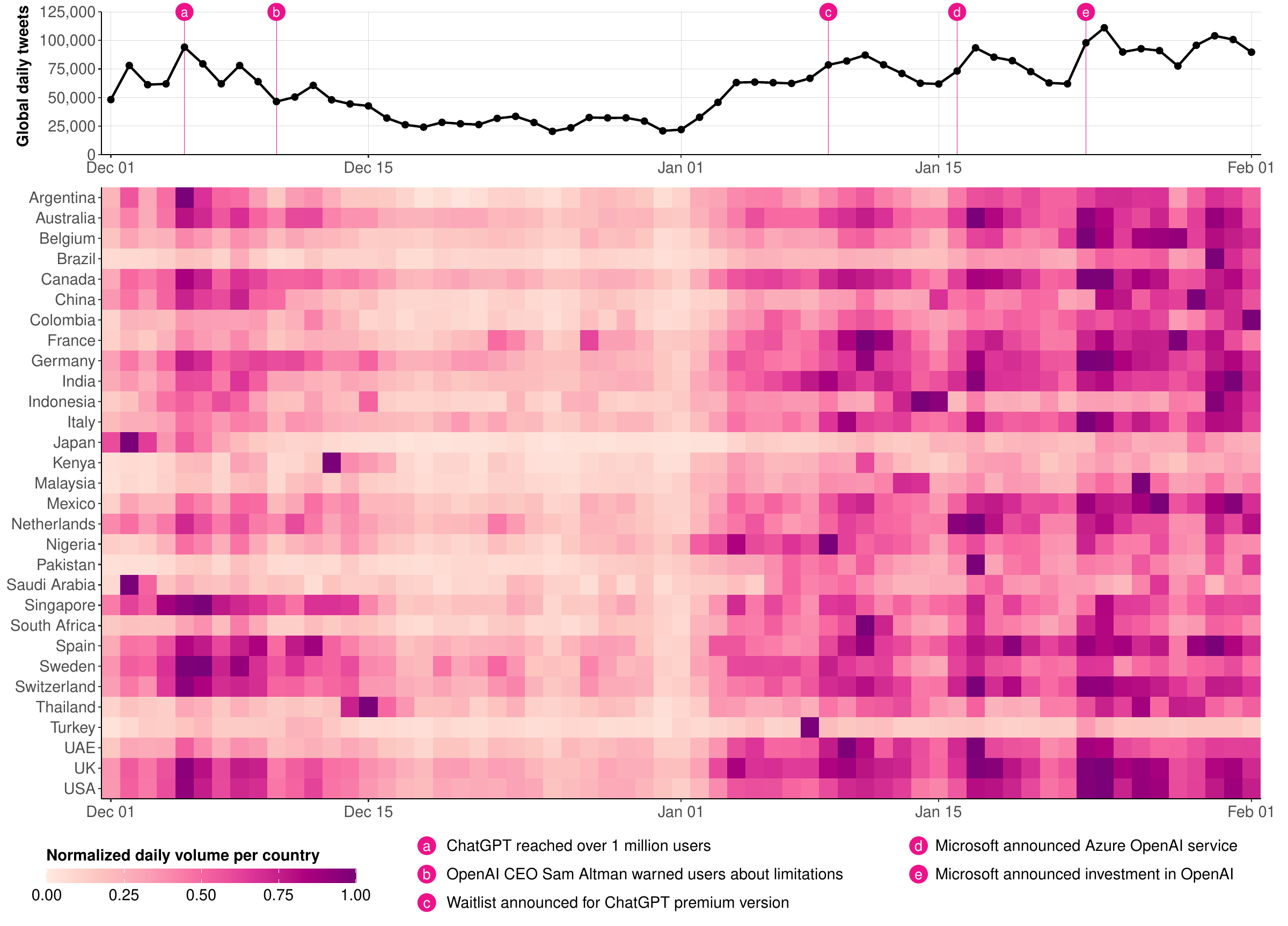}
\caption{Temporal distribution of tweets across countries. Only the 30 countries with the highest tweet volumes are shown.}
\label{fig:fig_03}
\end{figure}

Our findings also suggest that users from cultures characterized by high uncertainty avoidance were less likely to express positive sentiment toward ChatGPT (Coef. = -0.002; 95\% CI = [-0.004, -0.000]; p = 0.016). The hesitancy may come from concerns about ethical risks, job displacement, or the reliability of AI-generated content, all of which were apparent during the initial period following ChatGPT’s release, which is the focus of our study. While not necessarily associated with overt negativity, the lower likelihood of positive sentiment may reflect a more measured and reserved stance, which is consistent with cultural preferences for stability, structure, and gradual adoption.

\begin{figure}[!htb]
\centering
\includegraphics[width=\textwidth,height=\textheight,keepaspectratio]{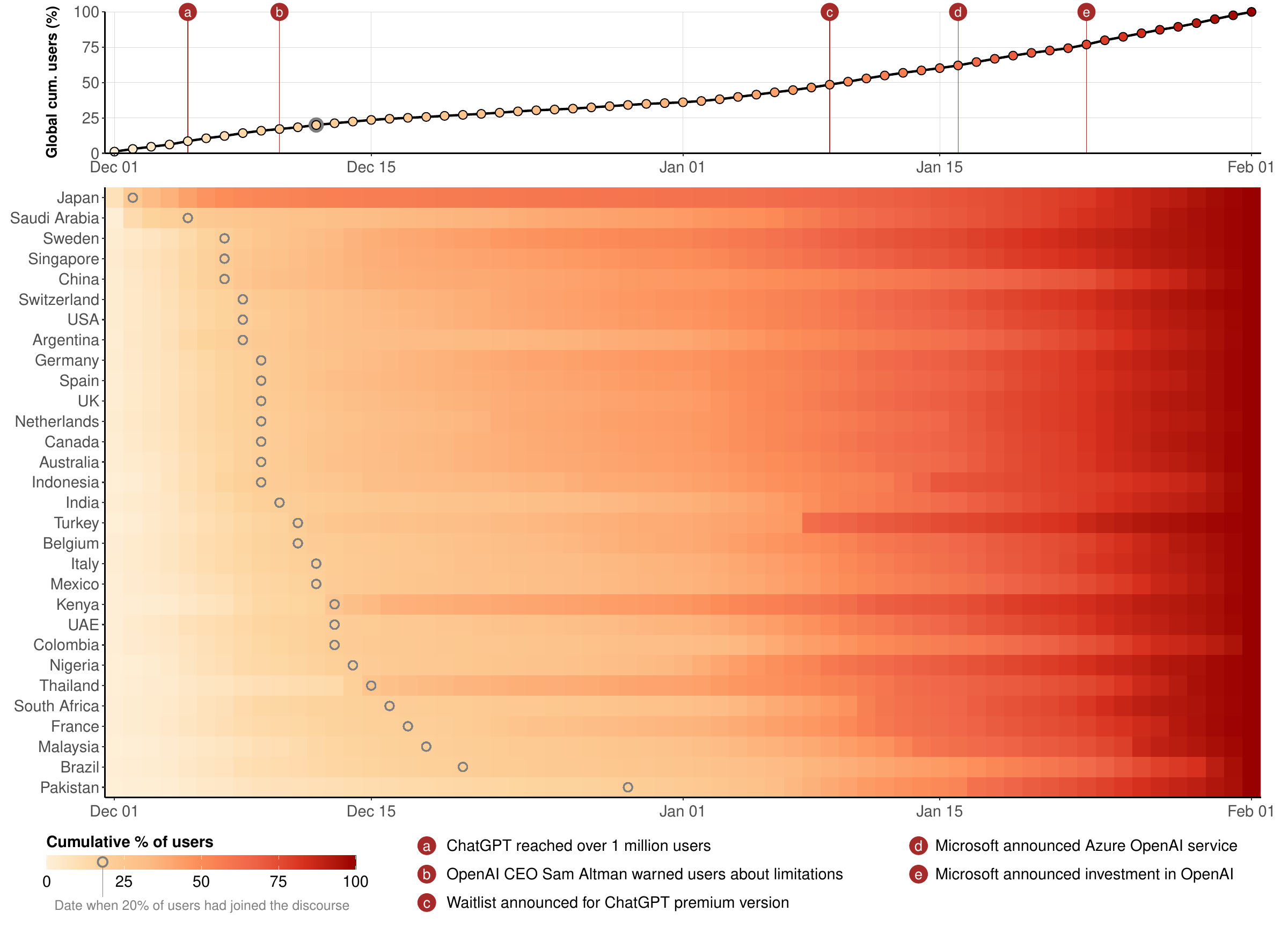}
\caption{Timing of initial engagement with ChatGPT discourse across countries. Only the 30 countries with the highest tweet volumes are shown. Countries are ordered by the date at which 20\% of users first engaged with the discourse.}
\label{fig:fig_04}
\end{figure}

In terms of timing (RQ4), our results indicate that users from more individualistic countries engaged earlier in the ChatGPT discourse (Coef = 0.003; 95\% CI = [0.002, 0.005]; p < 0.001), holding all other factors constant. European countries with high individualism scores, such as Sweden, Switzerland, Germany, the Netherlands, and the UK, were among the earliest to participate (see Figure \ref{fig:fig_04}). Japan (IDV = 62), traditionally collectivist but increasingly influenced by individualistic cultural norms through globalization, also exhibited early engagement. The majority of tweets from Japan were posted within the first five days following the launch, during which discussions focused on experimenting with ChatGPT’s features (Figure \ref{fig:fig_03}). In contrast, users from more collectivist countries such as Pakistan, Brazil, Malaysia, South Africa, Thailand, and Nigeria were among the late adopters.

Our empirical findings also found no significant effects of the power distance index on either attitudes (Positive vs. Neutral: Coef = 0.001; 95\% CI = [-0.001, 0.003]; p = 0.391; Negative vs. Neutral: Coef = -0.001; 95\% CI = [-0.003, 0.001]; p = 0.436) or the timing of users’ initial engagement (Coef = -0.001; 95\% CI = [-0.002, 0.001]; p = 0.393). Thus, our analysis shows that individualism was a consistent predictor in all models. Higher Uncertainty avoidance is only associated with a lower prevalence of a positive stance, and power distance did not show any association in our models.

Furthermore, we also conduct supplemental analyses using a multilevel linear regression with the average stance score per user as the continuous dependent variable (see Table \ref{tab:results_R7} of Supplementary Information). Our estimates reveal declining positivity among later adopters (Coef = -0.001; 95\% CI = [-0.002, -0.001]; p < 0.001), suggesting an evolving narrative and potentially a shift from initial enthusiasm toward more cautious evaluations as a broader and possibly more skeptical audience joins the discussion. This result supports the interpretation that the discussion about ChatGPT did not turn more critical because people were suddenly convinced about its negative aspects. It turned sour because the composition of interested speakers in the public negotiation of meaning toward this phenomenon shifted. 

\section{Discussion}
Our analysis of Twitter communication around the launch of ChatGPT as a focusing event reveals that both occupational background and the associated skills, as well as cultural factors, potentially influence perceptions and when people first showed interest in ChatGPT. 

Our results show that people with jobs that require higher writing skills are less likely to express a positive stance towards ChatGPT. At the same time, people with jobs that require higher mathematical skills, which are an indicator of more technical jobs, are more likely to express a positive stance towards ChatGPT. Our findings extend prior research from \cite{miyazaki_public_2023}. The difference in evaluation is not primarily one based on who is most exposed to AI, but rather shaped by a divide between more technical occupations and those that specifically work with text, such as creative occupations. The underlying reason might in some cases be a perceived threat of future replacement \cite{acemoglu_artificial_2022, frey_future_2017}, but it is also related to ethical reasons, which have been a factor identified in prior research for more negative sentiment \cite{miyazaki_public_2023}. 

In our analysis of when people first engaged with ChatGPT as a product on Twitter, our findings indicate that individuals with jobs that require programming skills were more likely to be active on Twitter earlier. In contrast, those with jobs that require high writing skills joined comparatively later. This is also evident in the descriptive results of occupations that had the fastest adoption rate in the Twitter discourse (see Figure \ref{fig:fig_02}), which shows that people with occupations that typically require higher programming skills are among the early adopters. These findings are also in line with prior Twitter studies that showed a stronger focus on technology early in the debates \cite{haque_i_2022} that later shifted to a more critical assessment \cite{fukuma_comparative_2024}. Our additional analyses also show that this shift in the debate can be primarily explained by the late adopters in the debate who have higher writing skills and are less likely to communicate a positive stance towards ChatGPT.

While some potential cultural factors have already been discussed in the literature as affecting attitudes towards AI \cite{oshaughnessy_what_2023, wilczek_government_2024}, we did not have strong expectations. The strongest predictor in our analysis was individualism. People in countries with high individualism were more likely to express their opinions early; however, they were also less likely to hold a positive stance and more likely to express an explicit negative stance towards ChatGPT. Our finding contrasts with prior survey research in the US context, which indicates that a higher level of individualism is associated with more positive attitudes towards AI \cite{oshaughnessy_what_2023}. Thus, our finding in the opposite direction, covering many countries instead of just a single country, suggests that future research should further investigate the role of individualism regarding AI in comparative studies.

A higher uncertainty avoidance only explained a lower probability that users expressed a positive stance towards ChatGPT. This is consistent with the idea that in cultures where ambiguity and unpredictability are met with discomfort, new technologies such as ChatGPT may be approached with greater caution. However, it did not explain the negative stance towards ChatGPT, nor did it explain the time of adoption. Two factors can explain the low explanatory power of uncertainty avoidance in our study. First, prior research has focused explicitly on the regulatory context and risk perceptions \cite{wilczek_government_2024}. However, this research has not considered potential benefits. Furthermore, when evaluating the risk and benefit perceptions of emerging technologies, these should be measured separately as individual dimensions, rather than relying on a one-dimensional score \cite{binder_measuring_2012}. And following that logic, our results are consistent, as people who may have higher risk perceptions (in a high uncertainty avoidance culture) are more cautious in their positive evaluations. Secondly, people from high uncertainty avoidance cultures have stronger privacy concerns regarding social media \cite{trepte_cross-cultural_2017} and are consequently less likely to express strong opinions on social media. Thus, future research should continue to include uncertainty avoidance as a predictor and rely on direct survey methods rather than social media data, as this would, if the variable introduced potential bias in our analysis, more accurately measure the opinions of people from high uncertainty-avoidance cultures.

Power distance, in contrast, did not predict the stance or whether people were early or later adopters. One plausible explanation for this is that PDI could influence engagement and reactions to ChatGPT in multiple, opposing ways. For example, users from high-PDI cultures might perceive ChatGPT positively as a means to reinforce established social hierarchies and authority structures, or negatively if the technology is viewed as disruptive to traditional authority structures. Similarly, users from low-PDI (egalitarian) countries might welcome ChatGPT’s potential to democratize access to knowledge and services (e.g., level the playing field in education). In contrast, others might criticize it for further increasing inequalities or challenging existing norms of fairness (i.e., favoring those with higher digital literacy or technology access). This suggests that the influence of PDI on public engagement with ChatGPT may be neutralized by such counteracting mechanisms, unlike the more precise and more direct effects associated with individualism (e.g., personal autonomy concerns) and uncertainty avoidance (e.g., risk aversion). Future research could specifically test these interpretations in survey studies, as egalitarianism has only been examined in the US context, where it was found to be a predictor of more positive attitudes towards AI \cite{oshaughnessy_what_2023}.

Taking a step back, our analysis reveals that digital platforms can serve as arenas for the collective negotiation of meaning in response to technological change, especially following significant events. This builds on prior research identifying Twitter as a space where people discuss political developments, such as televised debates or policy failures highlighted by major events \cite{jungherr_logic_2014, zhang_whose_2019}. Similarly, the launch of new technologies can trigger public reactions and debates.

These debates offer valuable insight into how people make sense of change, the roles they assume in public discourse, and the factors driving their engagement and positioning. When we account for economic self-interest and cultural predispositions, people respond to events in ways that reflect their pre-existing interests and preferences. Shifts in aggregate discourse are largely explained by changes in who participates, rather than changes in the positions of participants.

The public negotiation of meaning on Twitter is, therefore, less a site of opinion change and more a reflection of entrenched political divides, interests, and values. In this, we align with \citeauthor{gamson_media_1989}'s \citeyear{gamson_media_1989} argument that narrative persistence around technologies is shaped less by deliberation and more by the changing fortunes of the interests these narratives serve and their resonance with current events. Public sensemaking on Twitter, then, is best understood as an expression of social power and the perceived impact of technological change on existing power relations.

Our study also has several limitations. First, our empirical strategy and use of observational cross-sectional data means that we cannot establish causality. Throughout the paper, we have consistently interpreted our regression estimates as measures of association. Second, Twitter users are not representative of the general population, and social media platform penetration varies widely across countries (e.g., due to censorship) and occupations (e.g., underrepresentation of manual labor sectors relative to technology-related sectors). However, in our case, the biased representation presents an opportunity, as we were primarily interested in invested individuals and publics, which allows us a window into collective negotiations of meaning and public sensemaking \cite{jungherr_logic_2014}. Still, future research should employ survey methods to validate our findings with a more representative sample of the population. Third, we were unable to identify the occupations and countries of origin for all users. While the classifier generally worked well, some users do not provide any information in their Twitter profiles that allows for classification. Nevertheless, the consistency of interpretations observed across our robustness tests (models including only skill-related or cultural variables; see Supplementary Information) provides additional confidence in the reliability of our estimates.  

Our analysis reveals that when discussing attitudes towards AI, both occupational background and cultural factors play a role. Both aspects provide opportunities for future research, especially in the context of the understudied role of culture. Future research could also focus on more nuanced dynamics, such as how culture and job interact with each other when people evaluate AI. Another area for future research is an extended time frame that allows for evaluating whether the opinion formed during the focusing event has become stable. Especially with nascent technology, failures can quickly shift public perception. At the same time, we observed this on a smaller scale (e.g., the aforementioned case of the children’s book), as well as failures, such as in Google’s new launch of Gemini's image-generation feature \cite{heath_google_2024}, where the chatbot created historically inaccurate images. These widely covered events can shift attitudes toward AI, much like new product launches will in the future. Thus, the process of public negotiation of meaning is an ongoing process that warrants attention in future studies.

\section{Conclusion}
The impact of innovative technologies in general\textemdash and AI in particular\textemdash on the economy, politics, and society remains uncertain. However, focusing events offer moments in which people can exchange views, voice concerns, and connect with others who share similar perspectives, potentially laying the groundwork for future political mobilization. Digital media provide a valuable window into these processes, helping us better understand societal concerns and the emergence of new fault lines.

Importantly, these interactions do not need to be deliberative; they often serve as stages for expressing power relations and making claims and positions visible. While such dynamics are well understood in the context of classic focusing or media events, they should also be extended to include product launches, public prototypes, and CEO presentations. Public reactions to these events reveal how people imagine the future, where interests converge or conflict, and what tensions may emerge.

The digitally mediated negotiation of meaning thus offers a map of existing and anticipated societal cleavages\textemdash and a preview of possible futures. Crucially, these futures are not fixed; they remain open to collective shaping.

\section*{Acknowledgments}
\label{acknowledgements}
Adrian Rauchfleisch’s work was supported by the National Science and Technology Council, Taiwan (R.O.C) (Grant No 113-2628-H-002-018-).

\bibliographystyle{apacite}
\bibliography{chat}

\clearpage

\appendix
\appendix           
\part{}             
\parttoc            

\section{Classifiers}
\subsection{Stance towards ChatGPT}\label{app:SI_stance}
We first tested several off-the-shelf sentiment classifiers, but they did not reach an acceptable Cohen’s $\kappa$ when we tried to validate them. Another drawback was that these traditional approaches typically struggled with multilingual text classification. Prior research (Rathje et al., 2024) demonstrates that GPT models surpass other methods in multilingual classification tasks. Thus, we switched to OpenAI's \textit{gpt-4o-mini-2024-07-18} model with temperature set to 0 to ensure deterministic results. For quoted tweets, we appended '[QUOTED TWEET]' after the tweet text to clearly mark the quoted content. We used the following prompt:

"Classify the stance in the following tweet, where 'positive' means the tweet shows a favorable view towards ChatGPT and its impact, 'neutral' means the tweet doesn't show a particular preference or is objective about ChatGPT and its impact, and 'negative' means the tweet shows an unfavorable view or concerns towards ChatGPT and its impact. Respond only with 1='positive', 2='neutral', or 3='negative' [tweet text]”

We used the following system role: “You are a helpful research assistant who classifies the stance of tweets towards ChatGPT”

First, one author and two research assistants independently labeled 500 randomly selected tweets, reaching an acceptable intercoder reliability (Krippendorff’s $\alpha$ = 0.67). Based on this dataset, we developed and tested the prompt on an earlier OpenAI model before applying it to the latest version. After classifying all tweets, we drew a new random sample of 500 tweets for final validation by one author. The classification achieved a Cohen’s $\kappa$ = 0.78 (Overall accuracy=0.87; F1 positive = 0.87; neutral = 0.87; negative = 0.85). We used Google Translate to assist in the manual classification of non-English tweets. While a temperature of 0 ensures the most deterministic results, there are still, over time, potentially small differences. 

Note that there are 205 tweets that received an invalid classification due to some reply error; these entries were excluded from the regression models.

\subsection{Location identification}\label{app:SI_loc}
We first used OpenStreetMap to identify the location of users by just using the information from the location field for each account. While this approach gave us already a high accuracy (87.4\%), we tried in a second step OpenAI's \textit{gpt-4o-mini-2024-07-18} model with a temperature of 0 and the following prompt:

"Your task is to determine which country the given place name is located in.
If the name corresponds to an actual location (this can include a full country name, common abbreviation like 'D' for Germany, IATA airport codes also count as location, or an emoji like [flag emoji] for Qatar), return only the full country name (e.g., 'Germany', 'Qatar'). If there is more than one location mentioned, use the first one. If it does not represent a valid location, return 'NA'.
Here is the name of the place: [location information from user tweet metadata]"

We used the following system role: “You are a helpful assistant identifying what country a given place name is located”

We post-processed the results from the identification and manually cleaned systematic errors. The results were validated through a random sample of 500 users with identified locations and another 500 random sample of users with locations not identified (NA cases) using the method previously described. Manual validation results showed high accuracy for both users with identified locations (96.4\%) and those without identified locations (95.6\%). Furthermore, an in-depth validation was also conducted in which we holistically examined the profile of the users, beyond just the location field, to confirm their true country of origin. This achieved an accuracy of 82.65\% for 392 users with accessible profiles in the first quarter of 2025.

\subsection{Occupation classification}\label{app:SI_occ}
For the occupation classification based on the user descriptions, we used OpenAI's model \textit{gpt-4o-mini-2024-07-18} with a temperature of 0 and the following prompt:

"Classify the professional background of this individual user using the US Standard Occupational Classification (SOC). Use the format 'XX-XXXX.xx' for SOC codes or 'student: MAJOR' for students. If information is insufficient or if the user is not an individual human, respond with 'NA'. Do not provide any additional explanation or details. [user description]"

We used the following system role: “You are a helpful research assistant who classifies user descriptions.”

We added the part about students, as it is difficult to assign a specific occupation, which helped prevent false positives. We performed basic post-processing and manually adjusted all SOC codes that could not be matched (e.g., some codes refer to an older version of O*NET).

We conducted two sets of validation for the identified occupations of Twitter users. With a random sample of 200 unique users, two authors independently determined (1) whether the occupation (for 66 out of 200, an occupation could be identified) of the user can be inferred from the provided profile description or not, and then (2) whether the identified occupation is reasonable or not. We used this approach because the O*NET database contains 873 unique occupations. In 95.5\% of the cases, the two authors came to the same conclusion. For the first step (1), out of 200 users, there were very few false positive (Coder A=3; Coder B=4) and false negative (Coder A=3; Coder B=9) cases identified by the two authors. Even if we count each individual disagreement of at least one of the two coders with the classification (job can be identified/ not identified), 93\% of the classifications were evaluated as correct. For the second step (2), the overall accuracy was also high between the two coders (90.9\%). Both evaluated most of the ChatGPT job classifications as reasonable (89.4\%), given the information provided in the account description. Even if we count each individual disagreement of at least one of the two coders with the classification, 84.8\% of the classifications were evaluated as reasonable.    

\section{Descriptive statistics}
\begin{table}[H]
\resizebox{\textwidth}{!}{
\begin{tabular}{L{4cm}cccccc}
\toprule
Variable & $M$ (SD) & Median & Min & Max & Skewness & Kurtosis\\
\midrule
Time Elapsed (s) & 2,702,718.14 (1,756,139.04) & 3,027,784.50 & 15 & 5,447,692 & -0.07 & -1.47\\
Writing Level  & 3.83 (0.60) & 3.88 & 1 & 5.75 & 0.18 & -0.44\\
Programming Level  & 1.78 (1.54) & 1.00 & 0 & 4.88 & 1.07 & -0.35\\
Mathematics Level & 2.78 (0.85) & 2.88 & 0 & 6 & -0.02 & 0.90\\
PDI & 52.54 (17.77) & 40 & 11 & 100 & 0.47 & -0.92\\
IDV & 53.63 (23.17) & 60 & 0 & 100 & -0.78 & -0.20\\
UAI & 56.85 (19.22) & 48 & 8 & 100 & 0.60 & -0.79\\
\bottomrule
\end{tabular}
}
\caption{Descriptive statistics of the variables used in the multilevel Cox proportional hazards regression model. PDI = Power Distance Index; IDV = Individualism vs.\ Collectivism; UAI = Uncertainty Avoidance Index.}
\label{tab:descr_D1}
\end{table}

\begin{figure}[!htb]
\centering
\includegraphics[width=\textwidth,height=\textheight,keepaspectratio]{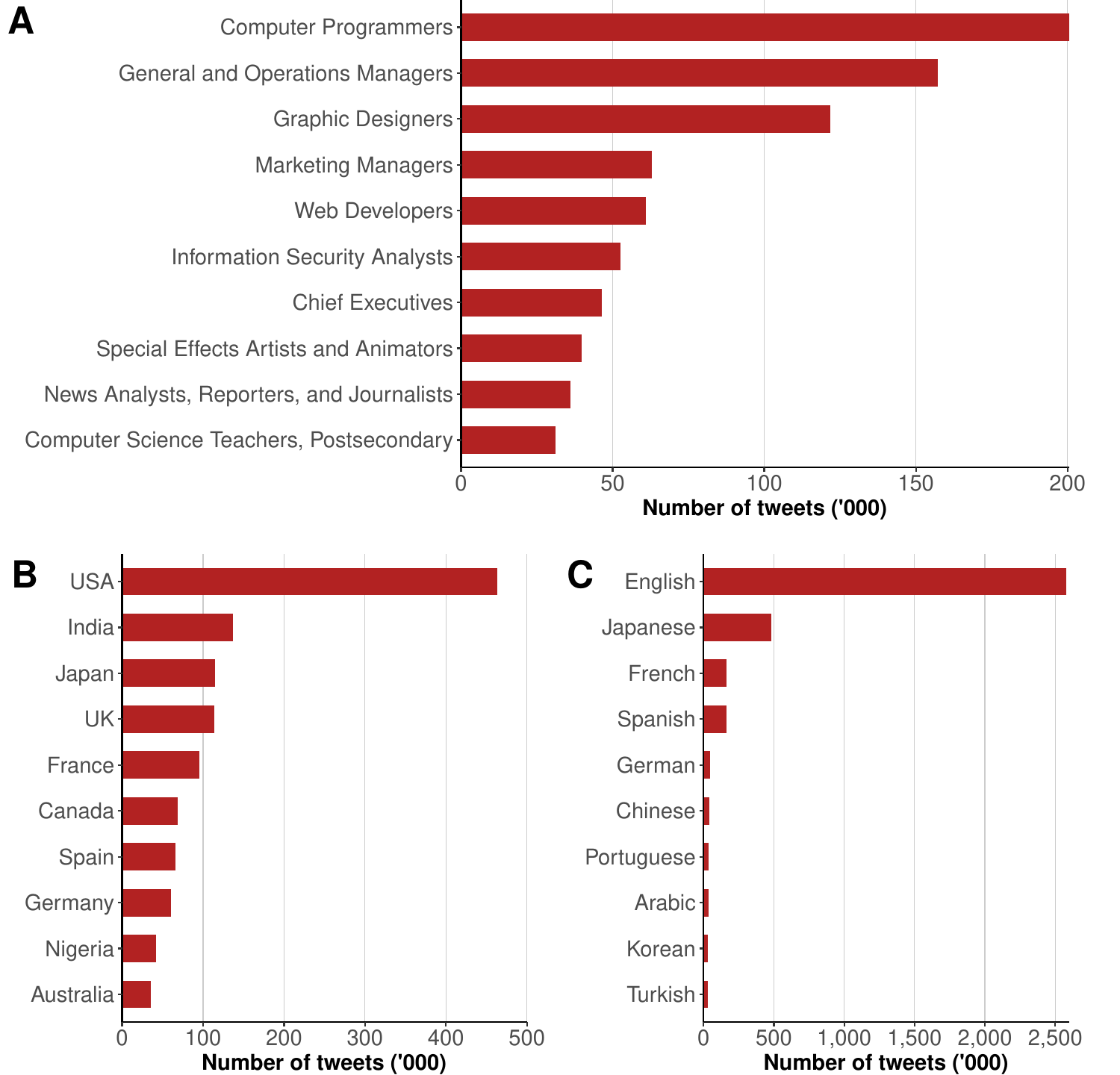}
\caption{Highest tweet volume by (A) user occupation, (B) user country, and (C) tweet language.}
\label{fig:fig_01_app}
\end{figure}

\begin{figure}[!htb]
\centering
\includegraphics[width=\textwidth,height=\textheight,keepaspectratio]{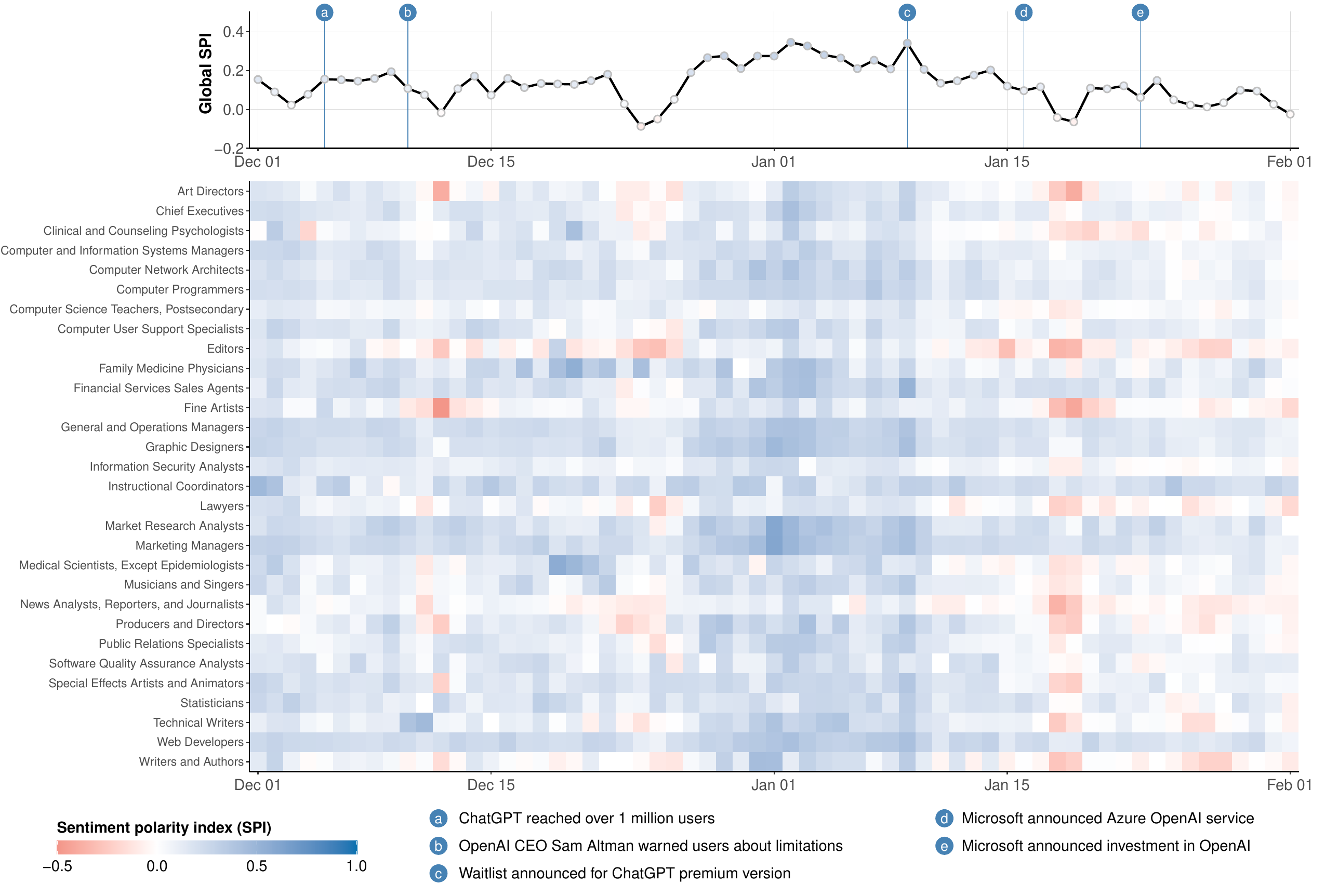}
\caption{Temporal changes of tweet sentiment across occupations. Only the 30 occupations with the highest tweet volumes are shown. SPI is calculated as the difference between the number of positive and negative tweets, divided by the total number of tweets.}
\label{fig:fig_02_app}
\end{figure}

\begin{figure}[!htb]
\centering
\includegraphics[width=\textwidth,height=\textheight,keepaspectratio]{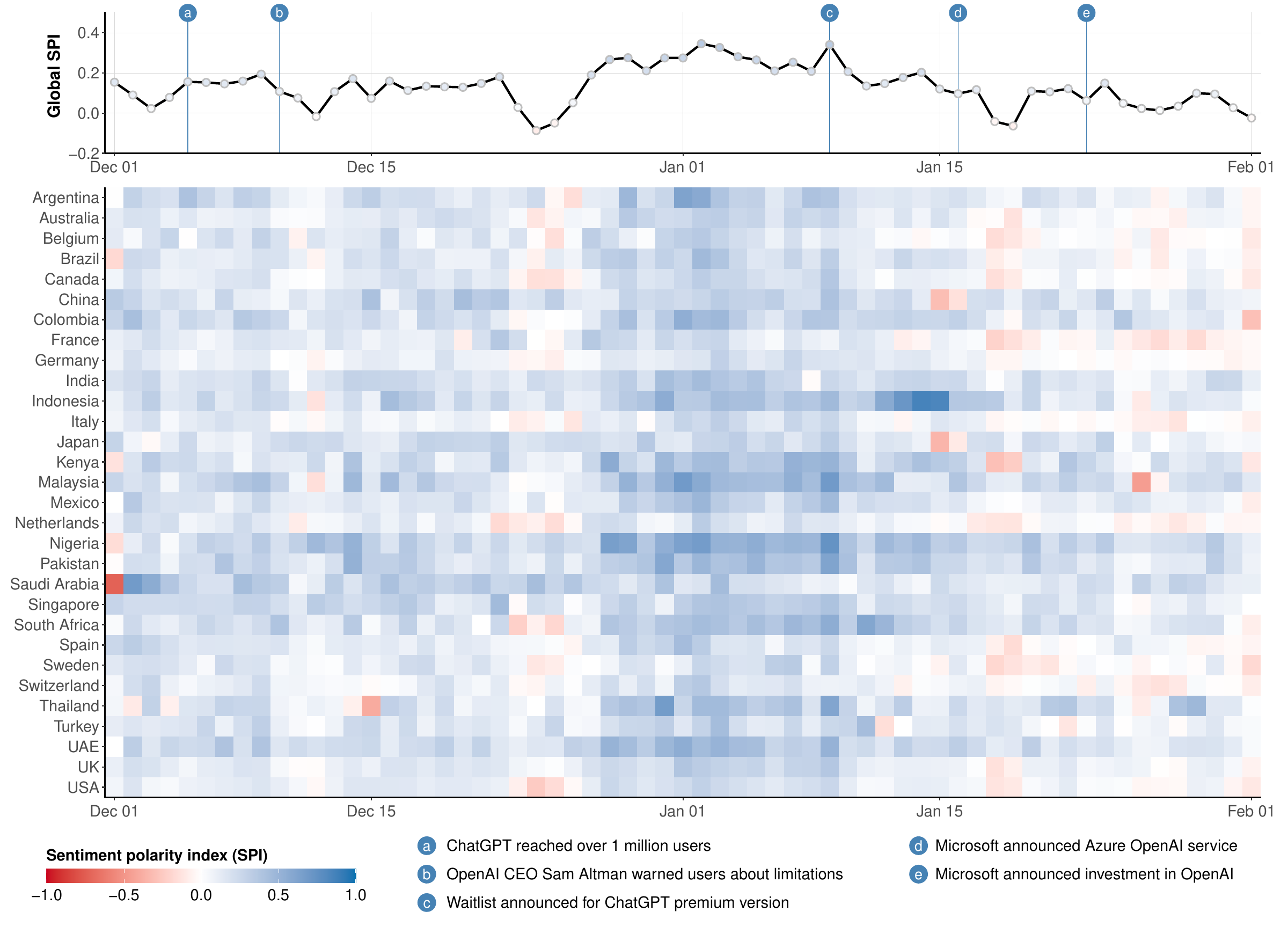}
\caption{Temporal changes of tweet sentiment across countries. Only the 30 countries with the highest tweet volumes are shown. SPI is calculated as the difference between the number of positive and negative tweets, divided by the total number of tweets.}
\label{fig:fig_03_app}
\end{figure}

\begin{table}[H]
\resizebox{\textwidth}{!}{
\begin{tabular}{L{4cm}cccccc}
\toprule
Variable & $M$ (SD) & Median & Min & Max & Skewness & Kurtosis\\
\midrule
Writing Level & 3.83 (0.59) & 3.88 & 1 & 5.75 & 0.26 & -0.43\\
Programming Level & 1.88 (1.58) & 1.25 & 0 & 4.88 & 0.93 & -0.67\\
Mathematics Level  & 2.80 (0.83) & 2.88 & 0 & 6 & 0.03 & 1.29\\
PDI & 52.01 (17.46) & 40 & 11 & 100 & 0.51 & -0.82\\
IDV & 54.78 (22.38) & 60 & 0 & 100 & -0.88 & 0.05\\
UAI & 57.12 (19.59) & 48 & 8 & 100 & 0.59 & -0.85\\
\bottomrule
\end{tabular}
}
\caption{Descriptive statistics of the variables used in the multilevel multinomial logistic model. PDI = Power Distance Index; IDV = Individualism vs.\ Collectivism; UAI = Uncertainty Avoidance Index.}
\label{tab:descr_D2}
\end{table}

\clearpage

\section{Main models}\label{app:SI_full}
In this section, we present the full multilevel regression results for both our stance (multinomial logit) model and our timing (Cox) model.  These tables include all coefficients discussed in the main text.
\begin{table}[H]
\resizebox{\textwidth}{!}{
\begin{tabular}{L{4cm}cccccc}
\toprule
  & \multicolumn{3}{c}{\textbf{Positive vs. Neutral}} & \multicolumn{3}{c}{\textbf{Negative vs. Neutral}}\\
\cmidrule(lr){2-4}\cmidrule(lr){5-7}
 & Coef.\ & 95\% CI & $p$ & Coef.\ & 95\% CI & $p$\\
\midrule
(Intercept)              & $-0.160$     & [$-0.418$, 0.098]     & 0.225     & $-1.287$  & [$-1.547$, $-1.027$] & $<0.001$\\
Writing Level            & $-0.104$     & [$-0.150$, $-0.059$]  & $<0.001$  & 0.002  & [$-0.047$, 0.050] & 0.943\\
Programming Level        &  $-0.017$    & [$-0.058$, 0.025]     & 0.425     & 0.009     & [$-0.035$, 0.053]       & 0.693\\
Mathematics Level        &  0.087       & [0.045, 0.128]        & $<0.001$   & $-0.060$     & [$-0.104$, $-0.015$]       & 0.008\\
PDI                      & 0.001        & [$-0.001$, 0.003]     & 0.391     & $-0.001$  & [$-0.003$, 0.001]    & 0.436\\
IDV                      &  $-0.003$    & [$-0.004$, $-0.001$]  & 0.005     & 0.006     & [0.004, 0.007]       & $<0.001$\\
UAI                      & $-0.002$     & [$-0.004$, $-0.000$]  & 0.016     & 0.001  & [$-0.000$, 0.003]        & 0.149\\
\midrule
\multicolumn{7}{l}{\textit{Random effects}}\\
Occupation & \multicolumn{6}{l}{\parbox[h]{\textwidth}{Positive vs. Neutral (Var = 0.066); Negative vs. Neutral (Var = 0.071);\\Positive and Negative (Cov = $-0.019$)}}\\
Country & \multicolumn{6}{l}{\parbox[h]{\textwidth}{Positive vs. Neutral (Var = 0.026); Negative vs. Neutral (Var = 0.020);\\Positive and Negative (Cov = 0.004)}}\\
\midrule
Observations ($N$)       & \multicolumn{6}{c}{878,264}\\
$N_{\text{occupation}}$  & \multicolumn{6}{c}{547}\\
$N_{\text{country}}$     & \multicolumn{6}{c}{117}\\
AIC                      & \multicolumn{6}{c}{1,641,015}\\
BIC                      & \multicolumn{6}{c}{1,641,247}\\
\bottomrule
\end{tabular}
}
\caption{Multilevel multinomial logit model estimates analyzing the factors affecting the stance toward ChatGPT with varying intercepts for occupation and country. Coef. = coefficient; AIC = Akaike Information Criterion; BIC = Bayesian Information Criterion; CI = confidence interval; PDI = Power Distance Index; IDV = Individualism vs.\ Collectivism; UAI = Uncertainty Avoidance Index.}
\label{tab:main_M1}
\end{table}

\begin{table}[H]
\resizebox{\textwidth}{!}{
\begin{tabular}{L{4cm}ccc}
\toprule
\textbf{Predictors} & Coef.\ & 95\% CI & $p$\\
\midrule
Writing Level            & $-0.027$     & [$-0.047$, $-0.006$]  & 0.010\\
Programming Level        &  0.095    & [0.078, 0.112]     & $<0.001$\\
Mathematics Level        &  0.014       & [$-0.004$, 0.031]        & 0.140\\
PDI                      &$-0.001$        & [$-0.002$, 0.001]     & 0.393\\
IDV                      &  0.003    & [0.002, 0.005]  & $<0.001$\\
UAI                      & $-0.001$     & [$-0.002$, 0]  & 0.057\\
\midrule
\multicolumn{4}{l}{\textit{Random effects}}\\
Occupation Intercept Var & \multicolumn{3}{c}{0.009}\\
Country Intercept Var & \multicolumn{3}{c}{0.016}\\
\midrule
Observations ($N$)       & \multicolumn{3}{c}{312,858}\\
$N_{\text{occupation}}$  & \multicolumn{3}{c}{547}\\
$N_{\text{country}}$     & \multicolumn{3}{c}{117}\\
AIC                      & \multicolumn{3}{c}{7,280,204}\\
BIC                      & \multicolumn{3}{c}{7,283,077}\\
\bottomrule
\end{tabular}
}
\caption{ Multilevel Cox proportional-hazards model with varying intercepts for occupation and country analyzing the factors affecting the timing of initial engagement with ChatGPT discourse (time elapsed until the first tweet after the launch in seconds). Coef. = coefficient; AIC = Akaike Information Criterion; BIC = Bayesian Information Criterion; CI = confidence interval; PDI = Power Distance Index; IDV = Individualism vs.\ Collectivism; UAI = Uncertainty Avoidance Index.}
\label{tab:main_M2}
\end{table}

\begin{figure}[htbp]
  \centering
  \includegraphics[width=0.9\textwidth]{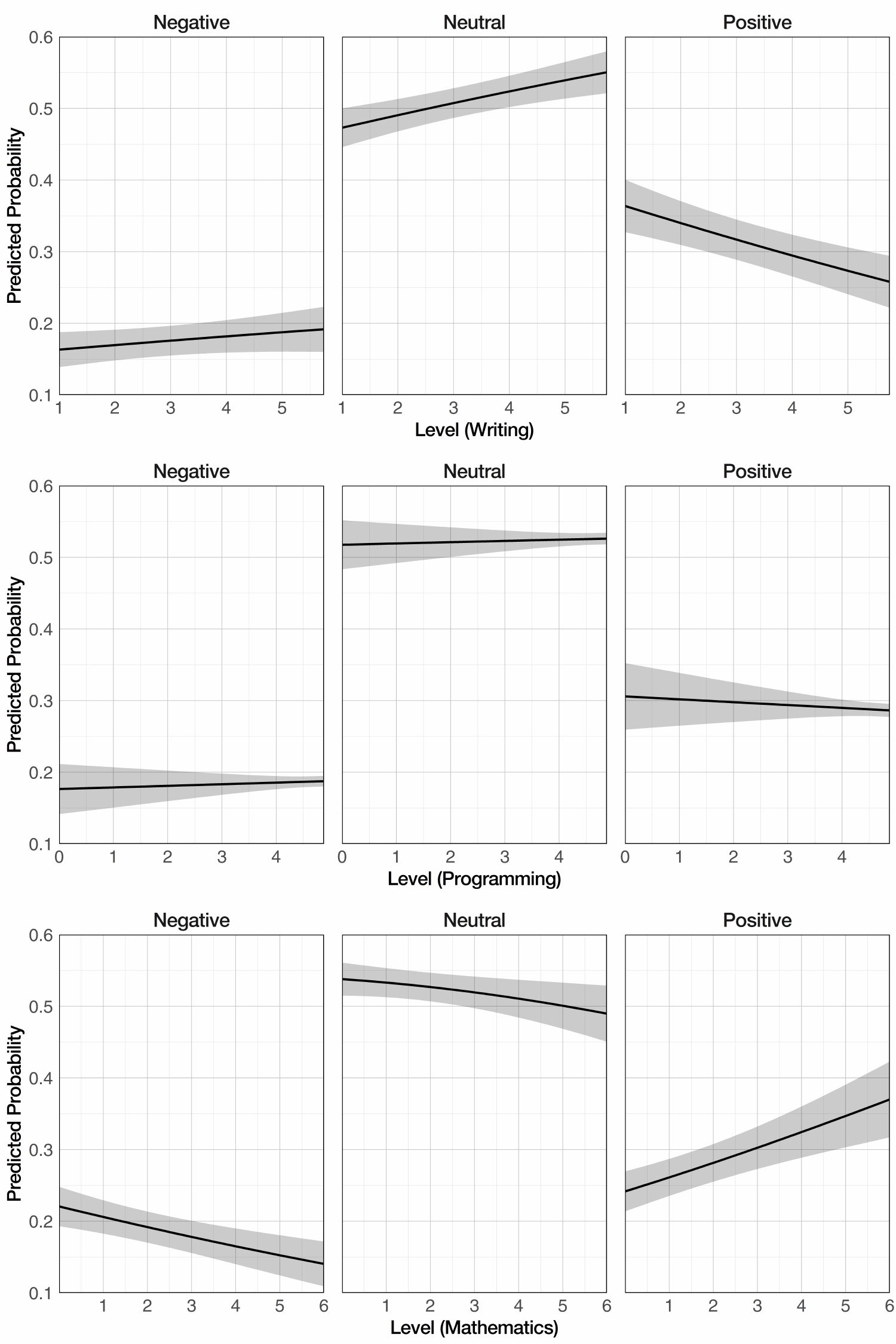}
  \caption{Conditional effects of occupation-related variables on the predicted probability of a tweet being negative, neutral, or positive.}
  \label{fig:main_MF1}
\end{figure}

\begin{figure}[htbp]
  \centering
  \includegraphics[width=0.9\textwidth]{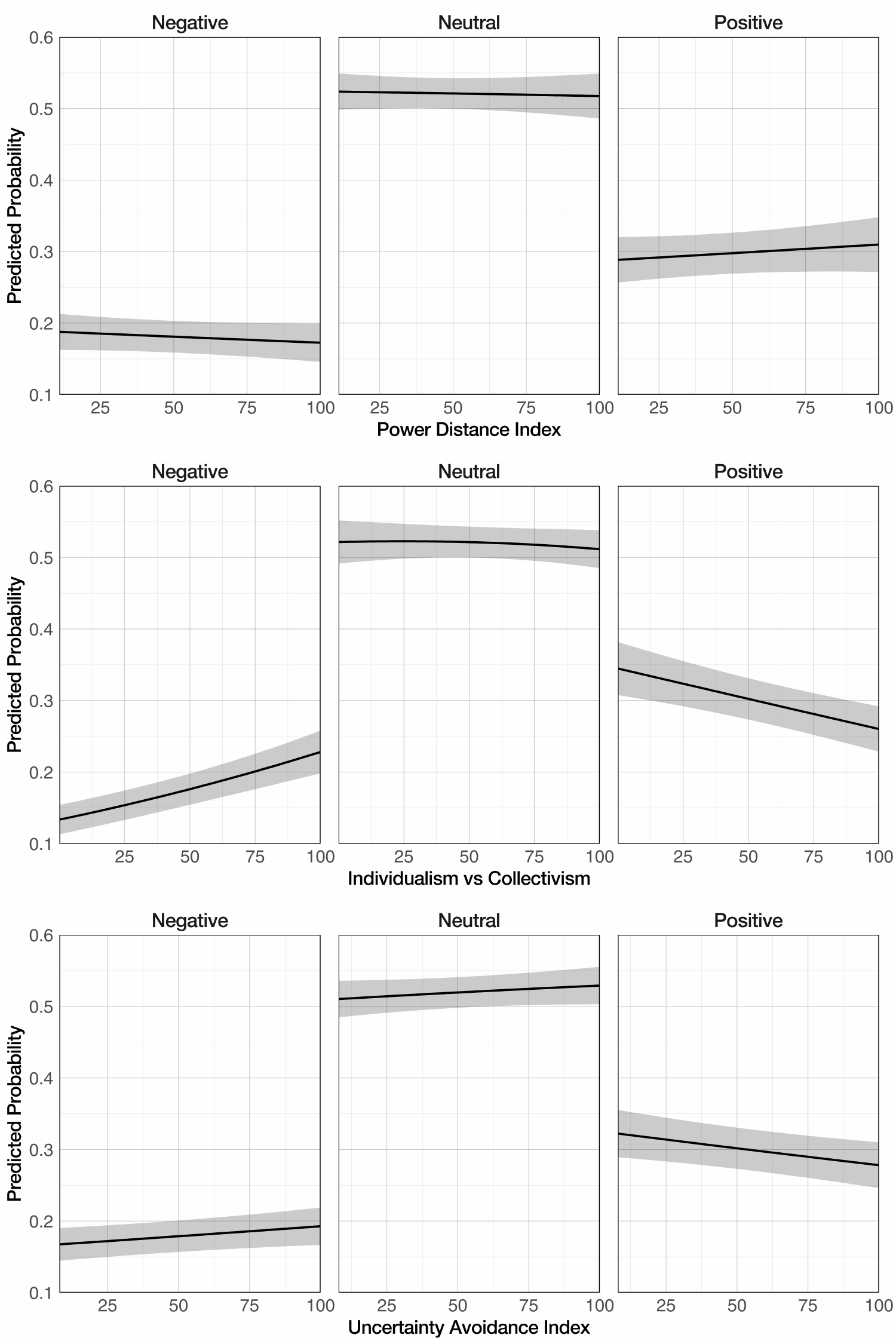}
  \caption{Conditional effects of country-related variables on the predicted probability of a tweet being negative, neutral, or positive.}
  \label{fig:main_MF2}
\end{figure}
\clearpage

\section{Robustness tests}\label{app:SI_rob}
We did a number of robustness tests. First, we checked whether our findings hold when we use the \textit{importance of the skill} required for an occupation instead of the \textit{level of skill} (see Table~\ref{tab:results_R1} for multilevel Cox regression model and Table~\ref{tab:results_R3}  for multilevel multinomial logit model). Second, we ran separate models using only occupation-related variables and only country-level variables (see Table~\ref{tab:results_R2} for multilevel Cox regression and Table~\ref{tab:results_R4} for multilevel multinomial logit). These subset models allowed us to include more observations (N), since the full model only includes observations with valid entries for both sets of variables. The consistency of the results across these subsets adds confidence to our main findings. Third, for modelling stance, we tested alternative specifications of the dependent variable: (a) treating stance as an ordered outcome using a multilevel ordinal logistic model (see Table~\ref{tab:results_R5}), and (b) using a multilevel linear model with the user’s average stance score as a continuous outcome (see Table~\ref{tab:results_R6}). Lastly, we included the date of the user’s initial engagement (i.e., the number of days since ChatGPT’s launch) as an additional independent variable in the multilevel linear model (Table~\ref{tab:results_R7}) to examine how sentiment may shift as new users join the conversation over time.

\begin{landscape}
\begin{table}[H]
\centering
\begin{tabular}{L{4cm}ccc@{\hspace{1cm}}ccc}
\toprule
  & \multicolumn{3}{c}{\textbf{Level model}} & \multicolumn{3}{c}{\textbf{Importance model}}\\
\cmidrule(lr){2-4}\cmidrule(lr){5-7}
 & Coef.\ & 95\% CI & $p$ & Coef.\ & 95\% CI & $p$\\
\midrule
Writing Skill            & $-0.027$ & [$-0.047$, $-0.006$] & 0.010  & $-0.049$ & [$-0.076$, $-0.022$] & $<0.001$\\
Programming Skill        &  0.095   & [0.078, 0.112]       & $<0.001$ & 0.151   & [0.124, 0.177]       & $<0.001$\\
Mathematics Skill        &  0.014   & [$-0.004$, 0.031]    & 0.140  & 0.030   & [0.005, 0.054]       & 0.018\\
PDI                      & $-0.001$ & [$-0.002$, 0.001]    & 0.393  & $-0.001$ & [$-0.002$, 0.001]    & 0.392\\
IDV                      &  0.003   & [0.002, 0.005]       & $<0.001$ & 0.003   & [0.002, 0.005]       & $<0.001$\\
UAI                      & $-0.001$ & [$-0.002$, 0]        & 0.057  & $-0.001$ & [$-0.002$, 0]        & 0.057\\
\midrule
\multicolumn{7}{l}{\textit{Random effects}}\\
Occupation Intercept SD  & 0.094 & Var = 0.009 &  & 0.093 & Var = 0.009 & \\
Country Intercept SD     & 0.125 & Var = 0.016 &  & 0.125 & Var = 0.016 & \\
\midrule
Observations ($N$)       & \multicolumn{3}{c}{312,858} & \multicolumn{3}{c}{312,858}\\
$N_{\text{occupation}}$  & \multicolumn{3}{c}{547}     & \multicolumn{3}{c}{547}\\
$N_{\text{country}}$     & \multicolumn{3}{c}{117}     & \multicolumn{3}{c}{117}\\
AIC                      & \multicolumn{3}{c}{7{,}280{,}204} & \multicolumn{3}{c}{7{,}280{,}202}\\
BIC                      & \multicolumn{3}{c}{7{,}283{,}077} & \multicolumn{3}{c}{7{,}283{,}066}\\
\bottomrule
\end{tabular}
\caption{Multilevel Cox proportional‐hazards models with varying intercepts for occupation and country predicting time to event (first tweet after the launch) from occupational skill levels and their perceived importance. PDI = Power Distance Index; IDV = Individualism vs.\ Collectivism; UAI = Uncertainty Avoidance Index.}
\label{tab:results_R1}
\end{table}
\end{landscape}

\begin{landscape}
\begin{table}[H]
\centering
\begin{tabular}{L{4cm}ccc@{\hspace{1cm}}ccc@{\hspace{1cm}}ccc}
\toprule
 & \multicolumn{3}{c}{\textbf{Full model}} & \multicolumn{3}{c}{\textbf{Occupation-only model}} & \multicolumn{3}{c}{\textbf{Country-only model}}\\
\cmidrule(lr){2-4}\cmidrule(lr){5-7}\cmidrule(lr){8-10}
 & Coef.\ & 95\% CI & $p$ & Coef.\ & 95\% CI & $p$ & Coef.\ & 95\% CI & $p$\\
\midrule
Writing Level         & $-0.027$ & [$-0.047$, $-0.006$] & 0.010   & $-0.019$ & [$-0.038$, $-0.001$] & 0.004 &   &   &  \\[2pt]
Programming Level     &  0.095   & [0.078, 0.112]       & $<0.001$ & 0.092   & [0.075, 0.109]       & $<0.001$ &   &   &  \\[2pt]
Mathematics Level     &  0.014   & [$-0.004$, 0.031]    & 0.140   & 0.007   & [$-0.010$, 0.025]    & 0.394 &   &   &  \\[2pt]
PDI                   & $-0.001$ & [$-0.002$, 0.001]    & 0.393   &         &                      &      & $-0.001$ & [$-0.003$, 0.001] & 0.162\\[2pt]
IDV                   &  0.003   & [0.002, 0.005]       & $<0.001$ &        &                      &      & 0.002 & [0, 0.004]         & 0.014\\[2pt]
UAI                   & $-0.001$ & [$-0.002$, 0]        & 0.057   &        &                      &      & $-0.001$ & [$-0.002$, 0]     & 0.101\\
\midrule
\multicolumn{10}{l}{\textit{Random effects}}\\
Occupation Intercept SD & 0.094 & Var = 0.009 & & 0.098 & Var = 0.010 & &   &   &  \\
Country Intercept SD    & 0.125 & Var = 0.016 & &     &              & & 0.123 & Var = 0.015 & \\
\midrule
Observations ($N$)      & \multicolumn{3}{c}{312,858} & \multicolumn{3}{c}{493,946} & \multicolumn{3}{c}{704,926}\\
$N_{\text{occupation}}$ & \multicolumn{3}{c}{547}     & \multicolumn{3}{c}{570}     & \multicolumn{3}{c}{—}\\
$N_{\text{country}}$    & \multicolumn{3}{c}{117}     & \multicolumn{3}{c}{—}       & \multicolumn{3}{c}{118}\\
AIC                     & \multicolumn{3}{c}{7{,}280{,}204} & \multicolumn{3}{c}{11{,}952{,}025} & \multicolumn{3}{c}{17{,}564{,}149}\\
BIC                     & \multicolumn{3}{c}{7{,}283{,}077} & \multicolumn{3}{c}{11{,}954{,}432} & \multicolumn{3}{c}{17{,}565{,}303}\\
\bottomrule
\end{tabular}
\caption{Multilevel Cox proportional‐hazards models comparing a full specification (varying intercepts for occupation and county) with occupation-only and country-only specifications. The outcome variable is the time elapsed until the first tweet after the launch in seconds. PDI = Power Distance Index; IDV = Individualism vs.\ Collectivism; UAI = Uncertainty Avoidance Index.}
\label{tab:results_R2}
\end{table}

\begin{table}[H]
\centering
\resizebox{19cm}{!}{%
\begin{tabular}{lccc@{\hspace{1cm}}ccc}
\toprule
\textbf{Positive vs. Neutral} & \multicolumn{3}{c}{\textbf{Level}} & \multicolumn{3}{c}{\textbf{Importance}} \\
\cmidrule(lr){2-4} \cmidrule(lr){5-7}
Predictor & Coef. & 95\% CI & $p$ & Coef. & 95\% CI & $p$ \\
\midrule
(Intercept)         & $-0.160$ & [$-0.418$, 0.098] & 0.225 & $-0.201$ & [$-0.504$, 0.102] & 0.194 \\
Writing Skill       & $-0.104$ & [$-0.150$, $-0.059$] & \textless0.001 & $-0.106$ & [$-0.167$, $-0.044$] & 0.001 \\
Programming Skill   & $-0.017$ & [$-0.058$, 0.025] & 0.425 & $-0.026$ & [$-0.092$, 0.040] & 0.438 \\
Mathematics Skill   & 0.087 & [0.045, 0.128] & \textless0.001 & 0.109 & [0.051, 0.166] & \textless0.001 \\
PDI                 & 0.001 & [$-0.001$, 0.003] & 0.391 & 0.001 & [$-0.001$, 0.003] & 0.406 \\
IDV                 & $-0.003$ & [$-0.004$, $-0.001$] & 0.005 & $-0.003$ & [$-0.005$, $-0.001$] & 0.007 \\
UAI                 & $-0.002$ & [$-0.004$, 0] & 0.016 & $-0.002$ & [$-0.004$, 0] & 0.019 \\
\midrule
Occupation Intercept Var & 0.066 & & & 0.068 & & \\
Country Intercept Var    & 0.026 & & & 0.028 & & \\
\midrule
\textbf{Negative vs. Neutral} & \multicolumn{3}{c}{\textbf{Level}} & \multicolumn{3}{c}{\textbf{Importance}} \\
\cmidrule(lr){2-4} \cmidrule(lr){5-7}
Predictor & Coef. & 95\% CI & $p$ & Coef. & 95\% CI & $p$ \\
\midrule
(Intercept)         & $-1.287$ & [$-1.547$, $-1.027$] & \textless0.001 & $-1.046$ & [$-1.366$, $-0.726$] & \textless0.001 \\
Writing Skill       & 0.002 & [$-0.047$, 0.050] & 0.943 & $-0.052$ & [$-0.118$, 0.013] & 0.117 \\
Programming Skill   & 0.009 & [$-0.035$, 0.053] & 0.693 & $-0.012$ & [$-0.081$, 0.057] & 0.738 \\
Mathematics Skill   & $-0.060$ & [$-0.104$, $-0.015$] & 0.008 & $-0.068$ & [$-0.129$, $-0.007$] & 0.028 \\
PDI                 & $-0.001$ & [$-0.003$, 0.001] & 0.436 & $-0.001$ & [$-0.003$, 0.001] & 0.493 \\
IDV                 & 0.006 & [0.004, 0.007] & \textless0.001 & 0.005 & [0.004, 0.007] & \textless0.001 \\
UAI                 & 0.001 & [0, 0.003] & 0.149 & $-0.001$ & [$-0.001$, 0.003] & 0.208 \\
\midrule
Occupation Intercept Cov/Var & $-0.019$ / 0.071 & & & $-0.020$ / 0.071 & & \\
Country Intercept Cov/Var    & 0.004 / 0.020 & & & 0.005 / 0.027 & & \\
\midrule
Observations ($N$)       & \multicolumn{3}{c}{878,264} & \multicolumn{3}{c}{878,264} \\
$N_{\text{occupation}}$  & \multicolumn{3}{c}{547}     & \multicolumn{3}{c}{547} \\
$N_{\text{country}}$     & \multicolumn{3}{c}{117}     & \multicolumn{3}{c}{117} \\
AIC                      & \multicolumn{3}{c}{1{,}728{,}808} & \multicolumn{3}{c}{1{,}728{,}809} \\
BIC                      & \multicolumn{3}{c}{1{,}729{,}042} & \multicolumn{3}{c}{1{,}729{,}043} \\
\bottomrule
\end{tabular}
}
\caption{Multilevel multinomial logistic regression with varying intercepts for occupation and country comparing positive and negative responses (vs.\ neutral) using occupational skill levels and their perceived importance as predictors. PDI = Power Distance Index; IDV = Individualism vs.\ Collectivism; UAI = Uncertainty Avoidance Index.}
\label{tab:results_R3}
\end{table}
\end{landscape}

\begin{landscape}
\begin{table}[H]
\centering
\resizebox{19cm}{!}{%
\begin{tabular}{lccc@{\hspace{0.8cm}}ccc@{\hspace{0.8cm}}ccc}
\toprule
\textbf{Positive vs. Neutral} & \multicolumn{3}{c}{\textbf{Full Level}} & \multicolumn{3}{c}{\textbf{Occupation Level}} & \multicolumn{3}{c}{\textbf{Country Level}} \\
\cmidrule(lr){2-4} \cmidrule(lr){5-7} \cmidrule(lr){8-10}
Predictor & Coef. & 95\% CI & $p$ & Coef. & 95\% CI & $p$ & Coef. & 95\% CI & $p$ \\
\midrule
(Intercept)         & -0.160 & [$-0.418$, 0.098] & 0.225   & -0.342 & [$-0.489$, $-0.195$] & \textless0.001 & -0.280 & [$-0.460$, $-0.101$] & 0.002 \\
Writing Level       & -0.104 & [$-0.150$, $-0.059$] & \textless0.001 & -0.104 & [$-0.147$, $-0.062$] & \textless0.001 &       &                         &      \\
Programming Level   & -0.017 & [$-0.058$, 0.025] & 0.425   & -0.028 & [$-0.069$, 0.013] & 0.183   &       &                         &      \\
Mathematics Level   & 0.087  & [0.045, 0.128]    & \textless0.001 & 0.087  & [0.047, 0.128]    & \textless0.001 &       &                         &      \\
PDI                 & 0.001  & [$-0.001$, 0.003] & 0.391   &        &                         &        & 0.001  & [$-0.001$, 0.002] & 0.550 \\
IDV                 & -0.003 & [$-0.004$, $-0.001$] & 0.005   &        &                         &        & -0.003 & [$-0.005$, $-0.001$] & 0.001 \\
UAI                 & -0.002 & [$-0.004$, 0]      & 0.016   &        &                         &        & -0.002 & [$-0.004$, $-0.001$] & 0.001 \\
\midrule
Occupation Intercept Var & 0.066 & & & 0.070 & & & & & \\
Country Intercept Var    & 0.026 & & &     & & & 0.023 & & \\
\midrule
\textbf{Negative vs. Neutral} & \multicolumn{3}{c}{\textbf{Full Level}} & \multicolumn{3}{c}{\textbf{Occupation Level}} & \multicolumn{3}{c}{\textbf{Country Level}} \\
\cmidrule(lr){2-4} \cmidrule(lr){5-7} \cmidrule(lr){8-10}
Predictor & Coef. & 95\% CI & $p$ & Coef. & 95\% CI & $p$ & Coef. & 95\% CI & $p$ \\
\midrule
(Intercept)         & -1.287 & [$-1.547$, $-1.027$] & \textless0.001 & -0.832 & [$-0.990$, $-0.674$] & \textless0.001 & -1.501 & [$-1.716$, $-1.286$] & \textless0.001 \\
Writing Level       & 0.002  & [$-0.047$, 0.050] & 0.943   & -0.003 & [$-0.048$, 0.042] & 0.906   &       &                         &      \\
Programming Level   & 0.009  & [$-0.035$, 0.053] & 0.693   & -0.015 & [$-0.059$, 0.028] & 0.484   &       &                         &      \\
Mathematics Level   & -0.060 & [$-0.104$, $-0.015$] & 0.008 & -0.066 & [$-0.109$, $-0.023$] & 0.003 &       &                         &      \\
PDI                 & -0.001 & [$-0.003$, 0.001] & 0.436   &        &                         &      & -0.001 & [$-0.004$, 0.001] & 0.268 \\
IDV                 & 0.006  & [0.004, 0.007]    & \textless0.001 &        &                         &      & 0.005  & [0.003, 0.007]    & \textless0.001 \\
UAI                 & 0.001  & [0, 0.003]        & 0.149   &        &                         &      & 0.002  & [0.001, 0.004]    & 0.006 \\
\midrule
Occupation Intercept Cov/Var & -0.019 / 0.071 & & & -0.018 / 0.074 & & & & & \\
Country Intercept Cov/Var    & 0.004 / 0.020  & & &                     & & & 0 / 0.032 & & \\
\midrule
Observations ($N$)     & \multicolumn{3}{c}{878,264} & \multicolumn{3}{c}{1,404,657} & \multicolumn{3}{c}{1,739,101} \\
$N_{\text{occupation}}$& \multicolumn{3}{c}{547}     & \multicolumn{3}{c}{570}       & \multicolumn{3}{c}{—} \\
$N_{\text{country}}$   & \multicolumn{3}{c}{117}     & \multicolumn{3}{c}{—}         & \multicolumn{3}{c}{118} \\
AIC                    & \multicolumn{3}{c}{1{,}728{,}808} & \multicolumn{3}{c}{2{,}791{,}581} & \multicolumn{3}{c}{3{,}476{,}414} \\
BIC                    & \multicolumn{3}{c}{1{,}729{,}042} & \multicolumn{3}{c}{2{,}791{,}715} & \multicolumn{3}{c}{3{,}476{,}551} \\
\bottomrule
\end{tabular}
}
\caption{Multilevel multinomial logistic regression using occupational skill levels and cultural dimensions across three model specifications (full, occupation-level only, and country-level only) for both positive and negative response contrasts. PDI = Power Distance Index; IDV = Individualism vs.\ Collectivism; UAI = Uncertainty Avoidance Index.}
\label{tab:results_R4}
\end{table}
\end{landscape}

\begin{landscape}
\begin{table}[H]
\centering
\resizebox{19cm}{!}{%
\begin{tabular}{lccc@{\hspace{1cm}}ccc}
\toprule
\textbf{Ordered Thresholds and Predictors} & \multicolumn{3}{c}{\textbf{Level}} & \multicolumn{3}{c}{\textbf{Importance}} \\
\cmidrule(lr){2-4} \cmidrule(lr){5-7}
 & Coef. & 95\% CI & $p$ & Coef. & 95\% CI & $p$ \\
\midrule
Threshold \texttt{-1\,|\,0}  & $-1.607$ & [$-1.668$, $-1.545$] & \textless0.001 & $-1.617$ & [$-1.685$, $-1.549$] & \textless0.001 \\
Threshold \texttt{0\,|\,1}   & 0.872    & [0.810, 0.933]       & \textless0.001 & 0.861    & [0.794, 0.929]       & \textless0.001 \\
Writing Skill       & $-0.047$ & [$-0.075$, $-0.018$] & 0.001   & $-0.022$ & [$-0.050$, 0.007]     & 0.142 \\
Programming Skill   & $-0.025$ & [$-0.099$, 0.048]    & 0.495   & $-0.012$ & [$-0.098$, 0.074]     & 0.777 \\
Mathematics Skill   & 0.082    & [0.045, 0.120]       & \textless0.001 & 0.070    & [0.033, 0.107]       & \textless0.001 \\
PDI                 & 0.021    & [$-0.010$, 0.052]    & 0.182   & 0.021    & [$-0.010$, 0.052]     & 0.183 \\
IDV                 & $-0.103$ & [$-0.137$, $-0.069$] & \textless0.001 & $-0.103$ & [$-0.137$, $-0.069$] & \textless0.001 \\
UAI                 & $-0.039$ & [$-0.065$, $-0.012$] & 0.004   & $-0.039$ & [$-0.065$, $-0.012$]  & 0.004 \\
\midrule
Occupation Intercept SD / Var & 0.296 / 0.088 & & & 0.301 / 0.090 & & \\
Country Intercept SD / Var    & 0.130 / 0.017 & & & 0.130 / 0.017 & & \\
\midrule
Observations ($N$)       & \multicolumn{3}{c}{878,264} & \multicolumn{3}{c}{878,264} \\
$N_{\text{occupation}}$  & \multicolumn{3}{c}{547}     & \multicolumn{3}{c}{547} \\
$N_{\text{country}}$     & \multicolumn{3}{c}{117}     & \multicolumn{3}{c}{117} \\
AIC                      & \multicolumn{3}{c}{1{,}732{,}483} & \multicolumn{3}{c}{1{,}732{,}490} \\
BIC                      & \multicolumn{3}{c}{1{,}732{,}600} & \multicolumn{3}{c}{1{,}732{,}607} \\
\bottomrule
\end{tabular}
}
\caption{Multilevel ordered logistic regression comparing the effects of occupational skill levels and their perceived importance with stance of the tweet (ordered: -1 = negative, 0 = neutral, 1 = positive). Thresholds reflect cut-points between ordinal categories. PDI = Power Distance Index; IDV = Individualism vs.\ Collectivism; UAI = Uncertainty Avoidance Index.}
\label{tab:results_R5}
\end{table}
\end{landscape}

\begin{landscape}
\begin{table}[H]
\centering
\resizebox{19cm}{!}{%
\begin{tabular}{lccc@{\hspace{1cm}}ccc}
\toprule
\textbf{Predictors} & \multicolumn{3}{c}{\textbf{Level}} & \multicolumn{3}{c}{\textbf{Importance}} \\
\cmidrule(lr){2-4} \cmidrule(lr){5-7}
 & Coef. & 95\% CI & $p$ & Coef. & 95\% CI & $p$ \\
\midrule
(Intercept)         & 0.215  & [0.134, 0.295]       & \textless0.001 & 0.154  & [0.059, 0.250]       & 0.002 \\
Writing Skill       & $-0.024$ & [$-0.039$, $-0.008$] & 0.003          & $-0.016$ & [$-0.037$, 0.005]   & 0.124 \\
Programming Skill   & $-0.004$ & [$-0.018$, 0.010]    & 0.542          & $-0.001$ & [$-0.024$, 0.021]   & 0.902 \\
Mathematics Skill   & 0.036  & [0.022, 0.050]       & \textless0.001 & 0.047  & [0.028, 0.067]       & \textless0.001 \\
PDI                 & 0      & [0, 0.001]           & 0.132          & 0      & [0, 0.001]           & 0.132 \\
IDV                 & $-0.001$ & [$-0.002$, $-0.001$] & \textless0.001 & $-0.001$ & [$-0.002$, $-0.001$] & \textless0.001 \\
UAI                 & $-0.001$ & [$-0.001$, 0]        & 0.005          & $-0.001$ & [$-0.001$, 0]        & 0.005 \\
\midrule
Occupation Intercept SD / Var & 0.084 / 0.007 & & & 0.085 / 0.007 & & \\
Country Intercept SD / Var    & 0.042 / 0.002 & & & 0.042 / 0.002 & & \\
\midrule
Observations ($N$)       & \multicolumn{3}{c}{312,847} & \multicolumn{3}{c}{312,847} \\
$N_{\text{occupation}}$  & \multicolumn{3}{c}{547}     & \multicolumn{3}{c}{547} \\
$N_{\text{country}}$     & \multicolumn{3}{c}{117}     & \multicolumn{3}{c}{117} \\
AIC                      & \multicolumn{3}{c}{563,265} & \multicolumn{3}{c}{563,265} \\
BIC                      & \multicolumn{3}{c}{563,371} & \multicolumn{3}{c}{563,372} \\
\bottomrule
\end{tabular}
}
\caption{Multilevel linear regression model with varying intercepts for occupation and country comparing the effects of occupational skill levels and their perceived importance with the outcome variable average stance score per user (continuous; based on individual tweet scores of -1 = negative, 0 = neutral, 1 = positive). PDI = Power Distance Index; IDV = Individualism vs.\ Collectivism; UAI = Uncertainty Avoidance Index.}
\label{tab:results_R6}
\end{table}
\end{landscape}

\begin{landscape}
\begin{table}[H]
\centering
\resizebox{19cm}{!}{%
\begin{tabular}{lccc@{\hspace{1cm}}ccc}
\toprule
\textbf{Predictors} & \multicolumn{3}{c}{\textbf{Level}} & \multicolumn{3}{c}{\textbf{Level + Days Elapsed}} \\
\cmidrule(lr){2-4} \cmidrule(lr){5-7}
 & Coef. & 95\% CI & $p$ & Coef. & 95\% CI & $p$ \\
\midrule
(Intercept)         & 0.215  & [0.134, 0.295]       & \textless0.001 & 0.266  & [0.186, 0.347]       & \textless0.001 \\
Writing Level       & $-0.024$ & [$-0.039$, $-0.008$] & 0.003          & $-0.022$ & [$-0.038$, $-0.007$] & 0.005 \\
Programming Level   & $-0.004$ & [$-0.018$, 0.010]    & 0.542          & $-0.007$ & [$-0.022$, 0.007]    & 0.306 \\
Mathematics Level   & 0.036  & [0.022, 0.050]       & \textless0.001 & 0.036  & [0.021, 0.050]       & \textless0.001 \\
PDI                 & 0      & [0, 0.001]           & 0.132          & 0.001  & [0, 0.001]           & 0.110 \\
IDV                 & $-0.001$ & [$-0.002$, $-0.001$] & \textless0.001 & $-0.002$ & [$-0.002$, $-0.001$] & \textless0.001 \\
UAI                 & $-0.001$ & [$-0.001$, 0]        & 0.005          & $-0.001$ & [$-0.001$, 0]        & 0.007 \\
Days Elapsed        & —      & —                    & —             & $-0.001$ & [$-0.002$, $-0.001$] & \textless0.001 \\
\midrule
Occupation Intercept SD / Var & 0.084 / 0.007 & & & 0.085 / 0.007 & & \\
Country Intercept SD / Var    & 0.042 / 0.002 & & & 0.042 / 0.002 & & \\
\midrule
Observations ($N$)       & \multicolumn{3}{c}{312,847} & \multicolumn{3}{c}{312,847} \\
$N_{\text{occupation}}$  & \multicolumn{3}{c}{547}     & \multicolumn{3}{c}{547} \\
$N_{\text{country}}$     & \multicolumn{3}{c}{117}     & \multicolumn{3}{c}{117} \\
AIC                      & \multicolumn{3}{c}{563,265} & \multicolumn{3}{c}{562,586} \\
BIC                      & \multicolumn{3}{c}{563,371} & \multicolumn{3}{c}{562,703} \\
\bottomrule
\end{tabular}
}
\caption{Multilevel linear model with varying intercepts for occupation and country and users' first engagement as additional independent variable. The outcome variable is the average stance score per user (continuous; based on individual tweet scores of -1 = negative, 0 = neutral, 1 = positive) PDI = Power Distance Index; IDV = Individualism vs.\ Collectivism; UAI = Uncertainty Avoidance Index.}
\label{tab:results_R7}
\end{table}
\end{landscape}

\section{Additional information}\label{app:SI_example}
These are some widely retweeted responses to the following tweet -- \textit{"I spent the weekend playing with ChatGPT, MidJourney, and other AI tools... and by combining all of them, published a children's book co-written and illustrated by AI! Here's how!"} Hashtags, links, media, and profane language have been removed or paraphrased for readability.

\begin{table}[H]
\resizebox{\textwidth}{!}{
\begin{tabular}{lL{12cm}L{4cm}l}
\toprule
\textbf{No} & \textbf{Tweet} & \textbf{Occupation} & \textbf{Country}\\
\midrule
1   & \parbox[h]{\textwidth}{As a children’s author/illustrator, it is saddening to see these books\\bc, apart from the ethics of AI and stolen artwork, kids deserve\\ better!!! I’m tired of people who see kidlit as an easy get-rich-\\quick scheme and putting in the absolute minimal effort into their\\books} & \parbox[t]{\textwidth}{Graphic Designers}  & NA\\
2   & \parbox[h]{\textwidth}{This is SO wrong. My sister recently illustrated a children’s novel\\for a big publisher for just around \$3000. 70 illustrations + cover\\art. What do you think this technology will do to human artists\\working in this field who are already super underpaid and\\overworked?}  & \parbox[t]{\textwidth}{Special Effects\\Artists and\\Animators}  & Germany\\
3   & \parbox[h]{\textwidth}{There’s something incredibly dystopian about teaching your kids\\“morality and life lessons” through a machine auto-generated\\book.}  & \parbox[t]{\textwidth}{Fine Artists,\\Including Painters,\\Sculptors, and\\Illustrators}  & Philippines\\
4   & \parbox[h]{\textwidth}{Children deserve better than badly cobbled together STOLEN art\\and words. This is such a soulless cash grab.}  & \parbox[t]{\textwidth}{Fine Artists,\\Including Painters,\\Sculptors, and\\Illustrators}  & NA\\
5   & \parbox[h]{\textwidth}{If you want to know what NOT to do with AI, this dude here ticks\\all the boxes. idea generated with AI \ding{51} took him a weekend \ding{51}\\claims the ownership of the whole thing \ding{51} poses proudly with his\\*biggest* accomplishment in life \ding{51} tries to make money out of it\\\ding{51}}  & Graphic Designers  & NA\\
6   & \parbox[h]{\textwidth}{Este señor ha publicado un libro ilustrado y escrito por IAs. Por\\favor, no compréis ni apoyéis nada de esto o acabaremos\\quedándonos sin cultura producida por humanos.\\(Translation: This man has published a book illustrated and written\\by AIs. Please do not buy or support any of this, or we’ll end up\\without any culture produced by humans.)}   & \parbox[t]{\textwidth}{Secondary School\\Teachers, Except\\Special and\\Career/Technical\\Education}  & Spain\\
7   & \parbox[h]{\textwidth}{I would never ever ever use AI for cover art or illustrations. I\\would never ever ever use AI to co-write a book!  This is\\disgusting and a slap in the face to all of us who do our own work\\and actually pay artists and photographers for book covers and\\illustrations.}  & NA  & NA\\
8   & \parbox[h]{\textwidth}{You didn’t make art, you didn’t co-write. You told machine to\\scrape thoughts and hardwork of living artists and writers,\\automated them trough machine learning because you didn’t want\\to pay them (cut costs! Innovate!) and then put a price tag on it\\to [satisfy] your vanity.}  & \parbox[t]{\textwidth}{Special Effects\\Artists and\\Animators}  & NA\\
9   & \parbox[!]{\textwidth}{AI: The laziest and stupidest way to make art.  It’s not even\\considered as art!}  & NA  & NA\\
10   & \parbox[h]{\textwidth}{There is no ‘debate’ to be had you [redacted], because in your head\\you’ll never see from an artists POV, never acknowledge how\\vile/unethical this stunt is, nor recognize the immorality. Artists\\from all backgrounds have voiced why they’re against this, and\\you fail to listen.}  & NA  & NA\\
\bottomrule
\end{tabular}
}
\caption{Public reactions to a tweet discussing the publication of a children's book generated using ChatGPT and other AI tools.}
\label{tab:addtl_A1}
\end{table}

\end{document}